\documentclass[aps,pra,twocolumn,floatfix]{revtex4-2}

\usepackage[thinc]{esdiff}
\usepackage{physics}

\usepackage{graphicx}% Include figure files
\usepackage{dcolumn}% Align table columns on decimal point
\usepackage{bm}% bold math
\usepackage{amsmath}
\usepackage{amssymb}
\usepackage{stackengine}

\begin{document}

%\preprint{APS/123-QED}

\title{Approaching the ultrastrong coupling regime between an Andreev level and a microwave resonator}

\author{O.O.~Shvetsov}
 \email{shvetsov@chalmers.se}
\author{A.~Khola}
\author{V.~Buccheri}
\author{I.P.C.~Cools}
\author{N. Trnjanin}
\affiliation{%
Department of Microtechnology and Nanoscience, Chalmers University of Technology, SE-41296 Gothenburg, Sweden
}
\author{T.~Kanne}
\author{J.~Nyg{\aa}rd}
\affiliation{
Niels Bohr Institute, Center for Quantum Devices, University of Copenhagen, Universitetsparken 5, DK-2100 Copenhagen O, Denmark
}

\author{A.~Geresdi}
\affiliation{%
Department of Microtechnology and Nanoscience, Chalmers University of Technology, SE-41296 Gothenburg, Sweden
}

\date{\today}% It is always \today, today,

\begin{abstract}

Josephson junctions formed in semiconductor nanowires host Andreev bound states and serve as a physical platform to realize Andreev qubits tuned by electrostatic gating. With the Andreev bound state being confined to the nanoscale weak link, it couples to a circuit-QED architecture via the state-dependent supercurrent flowing through the weak link. Thus, increasing this coupling strength is a crucial challenge for this architecture. Here, we demonstrate the fabrication and microwave characterization of a weak link which is defined in an InAs-Al (core-half shell) nanowire and embedded in a superconducting loop with a lumped-element resonator patterned from a thin NbTiN film with high kinetic inductance. We investigated several devices with various weak link lengths and performed spectroscopy that revealed pair transitions and single-quasiparticle transitions arising from spin-orbit split Andreev bound states. Our approach offers a compact geometry and a large resonator impedance above 12~k$\Omega$ at a resonator frequency of 8~GHz, which facilitates large coupling in the system. For the pair transitions, the experimentally observed energy level splitting demonstrates the coupling to an Andreev level of 490~MHz. We apply a perturbative model that shows good agreement with the experimental data and extract the maximum coupling of 968~MHz. Moreover, we show that the coupling is even stronger to an Andreev level with a higher transmission. In addition, spectroscopy of single-quasiparticle transitions reveals spin-orbit split Andreev bound states with the extracted spin-photon coupling of 77~MHz. 

\end{abstract}

\maketitle

\section{Introduction}

Circuit-QED (cQED) describes the interaction of photons stored in a superconducting resonator and a two-level system (qubit)~\cite{RevModPhys.93.025005}. The latter is often based on a single electron spin~\cite{Mi2016Dec} or collective excitations in various superconducting circuits~\cite{2008Natur.453.1031C, Wallraff2004Sep}. For quantum information processing, the favorable regime is when the qubit is strongly coupled to the resonator, that is, when they exchange a photon many times before the coherence is lost~\cite{Wallraff2004Sep}. 

A distinct ultrastrong coupling regime is established when the coupling strength becomes a significant fraction of the bare resonator and qubit frequencies~\cite{Forn-Diaz2019Jun, FriskKockum2019Jan, Qin2024Aug}. In this case, the common rotating-wave approximation is no longer valid and more comprehensive models are applied~\cite{Niemczyk2010Oct}. From a practical point of view, the ultrastrong coupling regime has been proposed to be used, e.g., for ultrafast quantum computation~\cite{Wang2017Mar}. 

Andreev bound states (ABSs) are discrete fermionic excitations spatially confined in a weak link between two superconductors~\cite{1969JETP...30..944K, PhysRevLett.66.3056}. Atomic-like transitions between these states can be utilized for the realization of two types of qubits: Andreev pair qubit~\cite{Desposito2001Sep,PhysRevLett.90.087003,Janvier2015Sep,PhysRevLett.121.047001, Zellekens2022Oct} and Andreev spin qubit~\cite{Chtchelkatchev2003Jun,Padurariu2010Apr,PhysRevX.9.011010,Hays2021Jul,Hays2020Nov,Bargerbos2023Aug,Pita-Vidal2023Aug}, which can be experimentally investigated by the means of cQED techniques~\cite{PhysRevResearch.3.013036,Matute-Canadas2022May,Wesdorp2024Jan}. Semiconductor nanowires serve as the preferred platform for such experiments, due to a high material quality, gate tunability, and intrinsically strong spin-orbit coupling~\cite{Krogstrup2015Apr}, which opens the way to exploit not only the charge, but also the spin degree of freedom in qubit applications. The possibility of coupling a single quasiparticle spin to macroscopic supercurrents makes Andreev qubits a promising alternative to traditional spin and superconducting qubit platforms~\cite{Padurariu2010Apr,Pita-Vidal2025Jan,Lu2024Dec}.

In conventional superconducting circuits, qubits are formed from quantized energy levels of collective electromagnetic excitations. In contrast, ABSs are localized within the weak link and carry a supercurrent up to the range of $10$~nA for the typical geometry of InAs nanowires ~\cite{PhysRevB.89.214508}. This makes it more challenging to reach a strong coupling to a microwave resonator. In practice, the coupling is typically realized by embedding a phase-biased weak link in a superconducting loop with a part of the resonator, also referred to as a shared inductance. In this case, the coupling rate $g_c$ is proportional to the zero-point flux (phase) fluctuations across the shared inductance~\cite{PhysRevResearch.3.013036}, which is in turn proportional to the square root of the resonator impedance $Z_r$, resulting in $g_c \propto \sqrt{Z_r}$~\cite{Devoret2007Oct, Stockklauser2017Mar}. Thus, an increase in the resonator impedance naturally leads to a stronger coupling, enabling better resolution spectroscopy and making higher energy states accessible.

In this paper, we investigate several devices that integrate a lumped-element resonator made of a thin NbTiN film and a weak link tailored in an InAs-Al (core-half shell) nanowire. Due to the high kinetic inductance of the NbTiN film, the differential impedance of the resonator reaches values above 12~k$\Omega$. We perform single- and two-tone spectroscopy that revealed pair and single-quasiparticle transitions. The system exhibits large energy splitting at the avoided crossings, which indicates the coupling to an Andreev level of 490~MHz and the model predicts that the coupling reaches up to 968~MHz. Moreover, we show that coupling to an Andreev level with a higher transmission can reach the value of 1.95~GHz, more than 20\% of the bare resonator frequency. Our findings suggest that the system is approaching the ultrastrong coupling regime. In addition, we demonstrate that the presented geometry can be used to resolve spin-orbit split ABSs, and we extract the coupling to an Andreev spin of 77~MHz.

\section{Device and setup}

\begin{figure}
\includegraphics[width=\columnwidth]{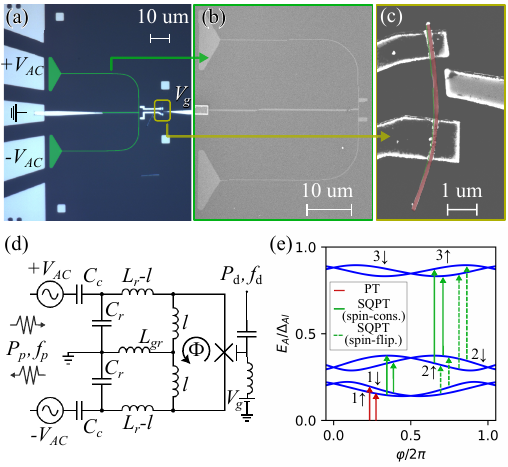}
\caption{(a) An optical micrograph of a representative device. The resonator is denoted by a green (false-color) U-shape structure patterned from a thin NbTiN film with high kinetic inductance. The resonator is capacitively coupled to a pair of feedlines. A weak link in an InAs-Al nanowire is defined close to the mirror line of the device and coupled to the resonator by aluminum contacts. (b) Electron micrograph of a test resonator structure fabricated simultaneously with the studied device on the same chip. (c) Electron micrograph of the nanowire (false-color) with a weak link, Al leads and a side gate. (d) Circuit diagram shows an equivalent lumped-element scheme of the device. It consists of a differential pair of resonators with the inductances $L_r$ and capacitances $C_r$, coupled to the feedlines by the capacitors $C_c$. The nanowire weak link is embedded in a flux-biased superconducting loop with a small fraction of the resonator with the shared inductance $2l$. To ground the loop, the resonator's middle part is connected with the ground plane. The resonator is probed in reflection with the probe signal of power $P_p$ and frequency $f_p$. The side gate is used to tune the nanowire's electrochemical potential and to excite the system with a drive signal of power $P_d$ and frequency $f_d$. (e) Typical excitation spectrum of a long weak link with spin-orbit interaction. Three ABSs are present in the spectrum, each of them is split into a doublet of states with the opposite pseudo-spin, denoted by $\uparrow$ and $\downarrow$. Microwave photons can induce pair transitions (PT, red arrows) and single-quasiparticle transitions (SQPT, green arrows), where for the latter spin-conserving (solid) and spin-flipping (dashed) processes can be distinguished. The energy scale is normalized to the superconducting gap in Al. }
\label{fig1}
\end{figure}

The devices used in this work consist of a weak link which is defined in an InAs nanowire~\cite{Krogstrup2015Apr} and embedded in a superconducting loop with a high-impedance lumped-element resonator. We report on fabrication and measurement of 3 devices, where the resonator geometry and the weak link length vary. Their parameters are listed in Table~\ref{tab:table1}.
    \begin{table*}%The best place to locate the table environment is directly after its first reference in text
    \caption{\label{tab:table1}%
    Parameters of the devices, including length, width and differential impedance of the resonators, thin film kinetic inductance, bare resonator frequency, zero-point flux fluctuations across the shared inductance and the weak link length.
    }
    \begin{ruledtabular}
    \begin{tabular}{llllllll}
    \textrm{Device}&
    \textrm{Length, $\mu$m\footnote{Length of one arm of the differential pair.}}&
    \textrm{Width, nm}&
    \textrm{$Z_{r,diff}$, k$\Omega$}&
    \textrm{$L_{k,\square}$, pH\footnote{May vary due to the sample ageing and slightly different fabrication parameters.}}&
    \textrm{$f_{r}$, GHz}&
    \textrm{$\Phi_{zpf}$, $\Phi_0$\footnote{Deduced from the resonator geometry and inductance.}}&
    \textrm{$l_{wl}$, nm}\\
    \colrule
    1 & 48.3 & 130 & 12.28 & 320 & 8.2230 & 0.016 & 250\\
    2 & 38.3 & 130 & 12.76 & 400 & 8.6190 & 0.021 & 320\\
    3 & 53.3 & 130 & 13.27 & 345 & 7.4685 & 0.015 & 450\\
    \end{tabular}
    \end{ruledtabular}
    \end{table*}
A representative device (replica of Device 2) is demonstrated in Fig.~\ref{fig1}(a) and is described below.

The resonator is patterned on a sapphire substrate with a thin (around 6~nm) sputtered film of a highly disordered NbTiN superconductor. Fig.~\ref{fig1}(b) represents an electron micrograph of the resonator structure. It consists of a differential pair of 130~nm wide and 38.3~$\mu$m long symmetric arms, divided by the grounding strip in the middle. Each arm is terminated with a triangular-shaped capacitor plate that is coupled to a 50~$\Omega$ feedline. When these feedlines are driven differentially, the microwave signal oscillates in the arms with the opposite phase, which is also referred to as the odd mode. In the even mode, on the contrary, microwaves in the two arms oscillate in-phase. Due to the high kinetic inductance of the thin film ${L_{k,\square} \approx 400}$~pH, this compact structure with a footprint of $50 \times 50~\mu $m$^{2} $ realizes a total inductance ${L_r \approx 118}$~nH and resonates in the odd mode at the frequency ${f_r = 8.619}$~GHz as a lumped-element resonator. The differential impedance of the resonator reaches ${Z_{r,diff} = 2\sqrt{\frac{L_r}{C_r}} \approx 12.76}$~k$\Omega$.

The weak link was defined in a 150~nm thick MBE-grown InAs nanowire with an in-situ grown epitaxial Al shell covering three of its facets~\cite{Krogstrup2015Apr}. A 250-450~nm long part of the 27 nm thick Al shell was removed by wet etching~\cite{Sestoft2024Jun}, an example is shown in the Fig.~\ref{fig1}(c). A small part of the resonator with the shared inductance $2l \approx$~12.3~nH is coupled to the nanowire by incorporating it in a common superconducting loop with Al contacts. The weak link couples to the odd mode of the resonator, because in the even mode the currents through the shared inductance cancel out. In this geometry, the coupling between the ABSs and the resonator is proportional to the zero-point flux fluctuations across the shared inductance ${\Phi_{zpf} = \frac{l}{L_r} \sqrt{\frac{\hbar Z_{r,diff}}{2}} \approx 0.021\Phi_0}$, where ${\Phi_0 = h/2e}$ is the superconducting flux quantum. The shared inductance was intentionally designed to be large, but comparable to the weak link Josephson inductance, which ensures that the phase across the junction $\varphi$ is a nonhysteric function of the applied flux $\Phi$. Moreover, our analysis shows that ${\varphi \approx 2\pi \Phi / \Phi_0}$ is a reasonable approximation (see Appendix D).

A DC electrostatic side gate is defined at the distance of 150~nm apart from the junction to tune the electrochemical potential in the nanowire. The differential drive and the overall symmetric chip layout ensure that along the mirror line, the AC electric field is essentially zero. To reference the nanowire to the DC ground potential, the nanowire loop is connected to the ground by a metal strip located along this line, which does not disturb the resonator's odd mode. In such a design, having the nanowire weak link defined at the zero AC voltage line is preferable to mitigate microwave losses through the side gate. A superconducting coil was used to flux bias the weak link. The circuit diagram of the device is presented in Fig.~\ref{fig1}(d).

The odd mode of the resonator is probed in reflection with the signal of power $P_p$ and frequency $f_p$. A separate microwave tone passes through a bias tee to the side gate to excite the system with a drive signal of power $P_d$ and frequency $f_d$.

After fabrication, the device chip is wire bonded to a printed circuit board and embedded in a copper box attached to the mixing chamber of a dilution refrigerator with a base temperature of 10~mK. To screen spurious magnetic fields in the laboratory, an additional aluminum shield wrapped with magnetic shielding foil was installed.

The system represents a typical cQED setup, where microwave photons drive transitions between ABSs, and the state of the system can be detected by the resonator dispersive shift. One of the defining parameters of the ABS spectrum in a weak link is the ratio between the weak link length $l_{wl}$ and the superconducting coherence length ${\xi}$, which is typically several hundred nanometers in InAs nanowires proximized by aluminum. To describe the excitation spectrum in a finite-length weak link with ${l_{wl}/\xi} \sim 1$, we refer to the model employed in Ref.~\cite{PhysRevX.9.011010,Park2017Sep}. The minimal model that includes spin-orbit splitting requires two transverse subbands, in this case the ABSs' energy dispersion ${\epsilon = E_A/\Delta_{Al}}$ is a solution to the transcendental equation:

\begin{align}
      &\tau \cos{[(\Lambda_1 - \Lambda_2)\epsilon \mp \varphi]} + (1-\tau)\cos{[(\Lambda_1 + \Lambda_2)\epsilon x_0]} =\nonumber\\
    &= \cos{[2\arccos{\epsilon}-(\Lambda_1 + \Lambda_2)\epsilon]},
\label{long_dispersion}
\end{align}

where $\tau$ is the channel transmission, $\Delta_{Al}$ is the Al superconducting gap, ${x_0 \in [-1,1]}$ indicates the position of a barrier in the weak link (normalized to ${l_{wl}/2}$), and ${\Lambda_{j} = l_{wl}\Delta_{Al}/\hbar v_{Fj}}$, where $v_{Fj}$ is the Fermi velocity of the subband ${j=1,2}$ in the nanowire.

In a finite-length weak link, multiple ABSs can squeeze into the spectrum, e.g., three levels are shown in a characteristic spectrum illustrated in Fig.~\ref{fig1}(e). Spin-orbit interaction lifts the spin degeneracy, splitting each of the states into a doublet with the opposite pseudo-spin, denoted by $\uparrow$ and $\downarrow$. The time-reversal symmetry preserves degeneracy only at ${\varphi = 0}$ and ${\varphi = \pi}$. We consider two types of transitions: a pair transition (PT, red arrows) takes place when two quasiparticles are excited from the ground state; a single-quasiparticle transition (SQPT, green arrows) occurs when a quasiparticle that is already present in the system is promoted to one of the upper levels. Among SQPTs, spin-conserving (solid) and spin-flipping (dashed) transitions can be distinguished, when the transition is between the levels with the same or opposite pseudo-spin, respectively.

\section{Results and discussion}

\begin{figure}
\includegraphics[width=\columnwidth]{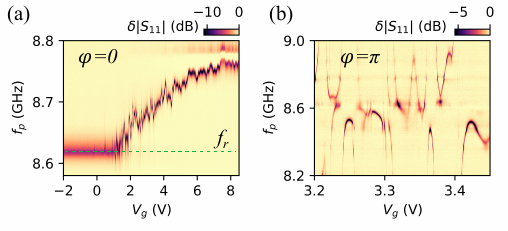}% Here is how to import EPS art
\caption{Change in the absolute value of the resonator reflection $\delta|S_{11}|$ as a function of the probe frequency $f_p$ and gate voltage $V_g$ measured in Device 2 at ${\varphi = 0}$ in (a) and at ${\varphi = \pi}$ in (b). The line at ${f_r=8.619}$~GHz in (a) denotes the bare resonator frequency, when the nanowire is fully pinched off. The gate-independent background was subtracted.}
\label{fig3}
\end{figure}

We first present the typical microwave gate response, obtained for Device 2. The change in the absolute value of the resonator reflection $\delta|{S_{11}}|$ as a function of the probe frequency $f_p$ and side-gate voltage $V_g$ is presented in Fig.~\ref{fig3} (the gate-independent background is subtracted). In (a), the phase across the weak link $\varphi$ is fixed at zero, while ${\varphi = \pi}$ in (b). The junction is fully pinched off below $V_g \approx$~1~V, where the resonance approaches the bare resonator frequency ${f_r = 8.619}$~GHz. The data obtained at ${\varphi = \pi}$ in (b) demonstrates multiple avoided crossings, signifying that the ABSs are strongly coupled to the resonator mode.

Further, we performed single- and two-tone spectroscopy to characterize the system. To mitigate charge noise and avoid electrostatic drifting, in these measurements we fix the gate voltage $V_g$ at one of the points where the resonator gate response has a local extremum, such as one of those observed in Fig.~\ref{fig3}(b). Single-tone spectra are taken by sweeping the probe frequency $f_p$ of power $P_p$ and measuring complex parameters $S_{11}$ at various applied fluxes. The flux-independent background was subtracted in the data plotted. When the two-tone spectra were measured, we first performed a single-tone frequency sweep to define the resonance position at a certain phase (flux). Then, a continuous wave probe tone was fixed at this frequency and $S_{11}$ was measured as a function of the drive tone with frequency $f_d$ and power $P_d$. To reference the signal measured in two-tone spectra, we subtracted a drive frequency-independent background.

\subsection{Spectroscopy of pair transitions}

\begin{figure}
\includegraphics[width=\columnwidth]{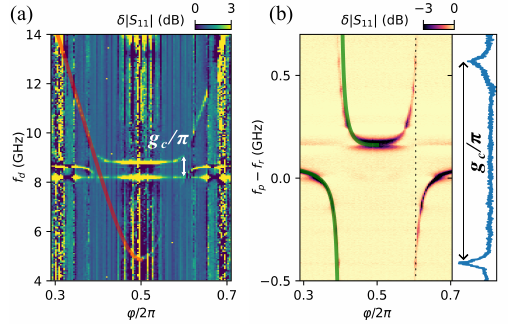}
\caption{(a) Two-tone spectrum measured in Device 2 at ${V_g = 3.39}$~V. The fit of the PT frequency as a function of phase is shown by the red line. The spectrum reveals avoided crossings at ${\varphi/2\pi=0.4}$ and ${\varphi/2\pi = 0.6}$ with the coupling at the crossings ${g_c/2\pi \approx 490}$~MHz, denoted by the white arrow. A flat line seen at around 8.1~GHz corresponds to the resonator's even mode, which is insensitive to the state of the weak link. (b) Corresponding single-tone spectrum. The probe frequency $f_p$ is offset to the bare resonator frequency ${f_r = 8.619}$~GHz. The green line represents the fit of the resonator shift. The line-cut on the right side clearly confirms ${g_c/2\pi \approx 490}$~MHz at the avoided crossing (along the gray dotted line in (b)).}
\label{fig4}
\end{figure}

Fig.~\ref{fig4} presents the spectroscopy data for Device 2. Data were acquired at ${V_g = 3.39}$~V and at the nominal powers ${P_p \approx -130}$~dBm and ${P_d \approx -100}$~dBm. The left panel illustrates the two-tone spectrum, which reveals a PT arising from a highly transmissive ABS. It crosses the odd mode of the resonator, forming a pair of avoided crossings at ${\varphi/2\pi = 0.4}$ and ${\varphi/2\pi = 0.6}$, where the coupling can be directly extracted from the level splitting on resonance ${g_c/2\pi \approx 490}$~MHz. 

In the vicinity of ${\varphi = \pi}$, the PT frequency can be fitted to the dispersion relation valid for a short weak link:

\begin{equation}
    hf_{PT} = 2E_A = 2\Delta'\sqrt{1-\tau \sin^2{(\varphi/2)}} 
    \label{short_dispersion}
\end{equation}

in respect that the effective gap value in a finite-length weak link is smaller than the gap in aluminum ${\Delta'<\Delta_{Al}}$ (see Appendix C for details). The least-squares method yields ${\Delta'/h=11.69\pm 0.03}$~GHz and ${\tau=0.9569\pm0.0005}$. The resulting fit is represented as a red line in Fig.~\ref{fig4}(a). 

Fig.~\ref{fig4}(b) shows the corresponding single-tone spectrum. The coupling at the avoided crossings ${g_c/2\pi \approx 490}$~MHz can be directly extracted from the line-cut along the gray dotted line, see the right side of the plot. The green line denotes the fit that was obtained from the model described in Appendix C. The fit relies on a single ABS observed in the two-tone spectrum in Fig.~\ref{fig4}(a). We confirmed that no higher-energy states are observed up to $f_d = 21$~GHz, see Appendix F. In the fit, the zero-point flux fluctuations across the shared inductance is the only fitting parameter. The least-squares fit to the experimental data yields ${\Phi_{zpf} = (0.0282\pm0.0002)\Phi_{0}}$, which slightly overshoots the value obtained from the geometry analysis, see Table~\ref{tab:table1}. To validate the fit, we extract the magnitude of the coupling at the avoided crossings ${g_c/2\pi=495\pm4}$~MHz, which is in good agreement with the observed energy splitting. The maximum coupling can be extracted at ${\varphi=\pi}$: ${g_c/2\pi=968\pm7}$~MHz.

\begin{figure}
\includegraphics[width=\columnwidth]{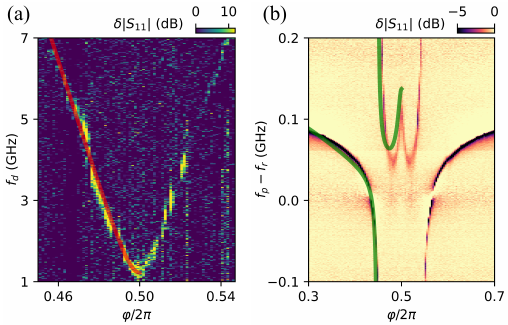}
\caption{(a) Two-tone spectrum measured in Device 1 at ${V_g = 14.5}$~V. The PT frequency fit plotted as a function of phase is denoted by the red line. (b) Corresponding single-tone spectrum. The probe frequency $f_p$ is offset to the bare resonator frequency ${f_r = 8.223}$~GHz. The green line denotes the fit of the resonator shift.}
\label{fig5}
\end{figure}

Fig.~\ref{fig5} provides the spectroscopy data, obtained for Device 1 with the shortest 250~nm weak link at ${P_p \approx -130}$~dBm and ${P_d \approx -100}$~dBm, and at ${V_g = 14.5}$~V (this device required a higher gate voltage due to the longer distance to the gate). Here, we were following a similar fitting routine. The two-tone spectrum in Fig.~\ref{fig5}(a) is fitted by Eq.~(\ref{short_dispersion}) describing a PT arising from a single high-transmission channel. The best fit reveals ${\tau=0.99942\pm0.00006}$ and ${\Delta'/h=25.2\pm0.2}$~GHz. The manifestation of ABS with a very large transmission in single-tone spectroscopy in Fig.~\ref{fig5}(b) is a sharp peak at ${\varphi = \pi}$~\cite{PhysRevResearch.3.013036}. This feature is well described within the presented model with ${\Phi_{zpf} = (0.0247\pm0.0002)\Phi_{0}}$ as the only fitting parameter, and we extract the maximum coupling ${g_c/2\pi=1.95\pm0.02}$~GHz at ${\varphi = \pi}$. We note that the extracted coupling indicates that the system is beyond the applicability of the perturbative approach. This explains a noticeable fit deviation in Fig.~\ref{fig5}(b). At this level of consideration, the extracted coupling should be perceived as an approximate value.

\subsection{Spectroscopy of single-quasiparticle transitions}

\begin{figure}[!htb]
\includegraphics[width=\columnwidth]{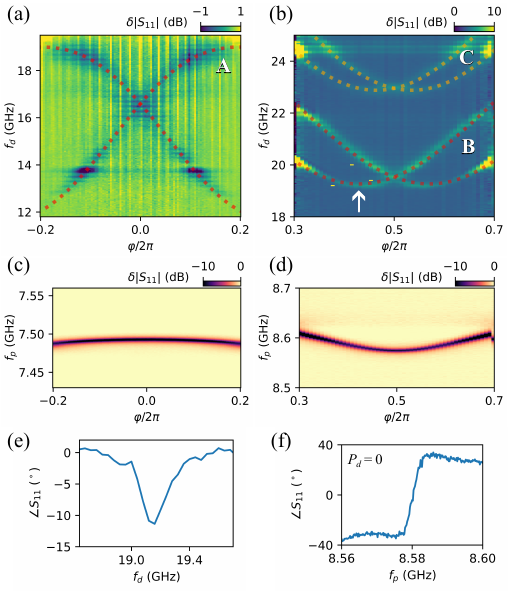}
    \caption{Spectroscopy data obtained for Device 3 at ${V_g = 1.498}$~V (a, c) and for Device 2 at ${V_g = 2.44}$~V (b, d). In two-tone spectra (a, b), bundles of SQPTs are observed, with crossings at the degeneracy points. SQPT frequency fits are denoted by the dotted lines and labeled as A, B and C. The fit parameters are listed in Table~\ref{tab:table2}. (c, d) Corresponding single-tone spectra. (e) Phase of the reflected signal as a function of the drive frequency at a minimum of the lower SQPT, see the white arrow in (b). (f) Corresponding resonator response when the system is not excited.}
\label{fig6}
\end{figure}

SQPTs associated with spin-orbit split ABSs were routinely observed for Devices 2 and 3 with longer weak links. 

Fig.~\ref{fig6}(a-d) provides the spectroscopy data obtained for Device 3 at ${V_g = 1.498}$~V and for Device 2 at ${V_g = 2.44}$~V around the degeneracy points ${\varphi = 0}$ and ${\varphi = \pi}$, respectively. Panels (a) and (b) illustrate two-tone spectroscopy data that reveal bundles of 4 lines, which we assign to spin-conserving SQPTs, see Fig.\ref{fig1}(c).

The spectroscopic lines in Fig.~\ref{fig6}(a) can be fitted with the transitions between the two lowest Andreev doublets ($1{\uparrow} \Rightarrow 2{\uparrow}$ and $1{\downarrow} \Rightarrow 2{\downarrow}$). Parameters which provide the best fit were obtained by the least-squares method applied to transitions between the ABSs given by Eq.~(\ref{long_dispersion}), see Appendix C for details. These parameters are listed in Table~\ref{tab:table2}, column A. The fit is shown by the red dotted line.
\begin{table}[b]%The best place to locate the table environment is directly after its first reference in text
\caption{\label{tab:table2}%
A table of parameters for the SQPT fits in Fig.~\ref{fig6}(a, b). $\Delta_{Al}/h$ is fixed at 52~GHz.
}
\begin{ruledtabular}
\begin{tabular}{cccc}
&
\textrm{A}&
\textrm{B}&
\textrm{C}\\
\colrule
$\Lambda_1$ & $5.7\pm0.3$ & $2.44\pm0.07$ & $2.8\pm0.1$\\
$\Lambda_2$ & $1.5\pm0.3$ & $3.06\pm0.09$ & $3.2\pm0.1$\\
$\tau$ & $0.35\pm0.04$ & $0.243\pm0.009$ & $0.265\pm0.006$\\
$x_0$ & $0.489\pm0.005$ & $0.2615\pm0.0001$ & -0.000837\\
\end{tabular}
\end{ruledtabular}
\end{table}

In Fig.~\ref{fig6}(b), two bundles of lines are observed. Each of those take place at relatively large frequencies above 19~GHz and have an upward curved shape, implying that the transitions occur between the upper doublets ($2{\uparrow} \Rightarrow 3{\uparrow}$ and $2{\downarrow} \Rightarrow 3{\downarrow}$). The parameters provided by the least-squares method are listed in Table~\ref{tab:table2}, where B corresponds to the red dotted line, and C to the orange one in Fig.~\ref{fig6}(b), respectively. 

The coupling between the resonator and a spinful ABS can be estimated from the line shape of a SQPT. Fig.~\ref{fig6}(e) shows the phase of the reflected signal as a function of the drive frequency at a
minimum of the lower SQPT in Fig.~\ref{fig6}(b) (marked by an arrow). The phase shift induced by this transition is ${\delta(\angle S_{11}) \approx 10^\circ}$. Based on the corresponding resonance line shape of a non-excited system (Fig.~\ref{fig6}(f)), the phase shift can be converted to the resonator frequency shift ${\delta f_r^{SQPT}\approx 770}$~kHz. It can be described by second-order perturbation theory \cite{Hays2020Nov}:

\begin{equation*}
      \delta f_r^{SQPT} \approx \left(\frac{g_c}{2\pi}\right)^2 \frac{2f_{SQPT}}{f_{SQPT}^2-f_r^2},
\end{equation*}

from which we extract the spin-photon coupling ${g_c/2\pi \approx 77}$~MHz. This value is comparable to the recently reported strong coupling between a high-impedance resonator and a singlet-triplet spin qubit~\cite{Ungerer2024Feb} and to the previously reported spin-photon coupling in InAs nanowire weak links~\cite{Hays2020Nov}. 

We note that a more rigorous assessment of the coupling requires measurements of parity-switching dynamics in the system, which is a subject for further investigation. If the non-equilibrium quasiparticles poisoning is weak, the system resides in an
odd-parity state~\cite{Fatemi2022Nov,PhysRevLett.121.047001,PhysRevResearch.3.013036,Sahu2024Apr} for only a small fraction of time, during which a SQPT can be excited. This can reduce the signal obtained in the spectroscopy of spin-orbit split ABSs.

In our experiment, we repeatedly observed that only spin-conserving SQPTs are clearly visible in the spectra. This is in agreement with the general selection rules: in the presence of transverse symmetry in the nanowire, spin-flipping transitions should be suppressed. However, previous studies showed that spin-flipping transitions can still be excited at elevated drive powers in a realistic non-ideal device~\cite{Hays2020Nov,PhysRevX.9.011010}. For example, the gate drive can violate transverse symmetry and unlock spin-flipping transitions~\cite{METZGER:2022iuj,PhysRevResearch.3.013036,Lu2025Jan}. Although we used the gate drive in our setup, we did not observe spin-flipping SQPTs. This is most likely because we had to operate at sufficiently low drive powers, so that the thin film resonator is not driven into a non-linear regime.

\section{Conclusion}
In conclusion, we demonstrated a robust approach for achieving a large coupling strength between a high-impedance lumped-element resonator and ABSs residing in an InAs nanowire weak link. Moreover, we showed that our geometry is suitable for the investigation of spin-orbit split ABSs and the associated Andreev spin qubits. The presented approach can be further extended to facilitate a strong qubit-qubit coupling in a device comprising multiple Andreev qubits.

\section{Data availability}

Raw datasets are available at the
Zenodo repository~\cite{shvetsov_2025_17117780}.

\section{Acknowledgments} 
This work was financially supported by the European Union's H2020 research and innovation program, grants \textnumero 804988 (SiMS) and \textnumero 828948 (AndQC) and by the Army Research Office (ARO) grant W911NF2210053. The Center for Quantum Devices is supported by the Danish National Research Foundation grant \textnumero DNRF101, the Novo Nordisk Foundation project SolidQ and the Carlsberg Foundation. The devices were fabricated in the Myfab cleanroom at Chalmers, and technical support from Nanofabrication Laboratory is gratefully acknowledged. We also thank P.~Makk, S.~Csonka, G.~F\stackon[1pt]{u}{..}l\stackon[1pt]{o}{..}p and V.~Fatemi for fruitful discussions. 

\appendix

\section{NbTiN thin film high kinetic inductance and the resonator impedance}

DC transport characterization of a test resonator structure (a replica of Device 1) provides NbTiN thin film normal state sheet resistance ${R_{N,\square} = 1.46}$~k$\Omega$ and critical temperature ${T_c = 6}$~K. With these parameters, we can estimate the sheet kinetic inductance as ${L^{DC}_{k,\square} \approx \frac{\hbar R_N}{\pi \Delta}}$~\cite{Annunziata2010Oct}, where $\Delta \approx 1.76k_B T_c$ is the superconducting gap linked to $T_c$ by a standard BCS prediction~\cite{PhysRev.108.1175}. This yields ${L^{DC}_{k,\square} \approx 320}$~pH.

In addition, for each of the devices, we performed microwave domain simulations for the actual geometries used in the experiment. To simulate a bare resonator, we replaced the weak links with a short break in the film, but all the aluminum structures were included because they contribute to the total stray capacitance of the resonator. We extracted $L_{k,\square}$ from the best fits of the simulated resonator microwave response to the experimental data at low $V_g$, when the nanowires are fully depleted. The results are listed in Table~\ref{tab:table1}, see the main text. Since all resonators were picked up from the same batch with the same NbTiN thickness, we attribute the variation in $L_{k,\square}$ to the aging of the sample and slightly different fabrication parameters used during nanowire processing (e.g., it was observed that the resulting $L_{k,\square}$ depends on the baking temperature of the resist).

The total inductance $L_r$ (for one arm of the differential pair) is derived from the geometry and $L_{k,\square}$ is obtained as described above. The bare resonator frequency $f_r$ is obtained experimentally. The differential impedance is defined as ${Z_{r,diff} = 2\sqrt{\frac{L_r}{C_r}} = 4\pi f_r L_r}$.

We note that we routinely observed the resonator's even mode, which is insensitive to the weak link's state and positioned at several hundred MHz below the odd mode, yielding a flat line at around 8.1~GHz in Fig.~\ref{fig4}. We confirmed the even mode frequency by microwave domain simulations. Its frequency is determined by ${2\pi f_r^{even} = 1/\sqrt{(L_r+2L_{gr})C_r}}$, where ${L_{gr} \approx 10}$~nH is the inductance of the grounding strip, which connects the ground plane and the middle part of the resonator, see Fig.~\ref{fig1}(a, d). 

\section{Measurement setup}

\begin{figure}
\includegraphics[width=\columnwidth]{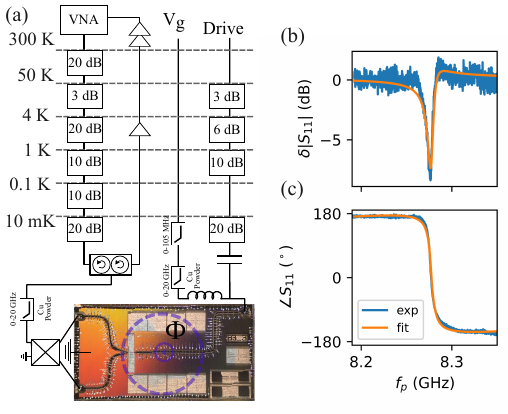}% Here is how to import EPS art
\caption{(a) Experimental setup. (b, c) Reflected signal amplitude (phase) versus probe tone frequency for one of the devices at zero external flux (the ABSs are far detuned). The blue line shows the experimental data. The orange line represents the fit, revealing ${Q_i \approx 3800}$ and ${Q_c \approx 1000}$.}
\label{fig2}
\end{figure}

The measurement scheme is similar to the one described in Ref.~\cite{Cools2025Jun} and depicted in Fig.~\ref{fig2}(a). The microwave readout tone passes through an attenuated line and a double circulator located at the mixing chamber stage, and then routed through a 180$^{\circ}$ hybrid to differentially drive two on-chip 50~$\Omega$ feedlines capacitively coupled to the resonator. The readout tone is reflected off the resonator's odd mode and routed through a series of cryogenic and room temperature amplifiers. The complex reflection parameter ($S_{11}$) is measured by a vector network analyzer. A separate microwave tone pass through the gate line to drive the ABS transitions, where a bias tee is used to combine the DC gate voltage and the drive tone. Fig.~\ref{fig2}(b, c) represents the characteristic $S_{11}$ response of one of the devices when the ABSs are far detuned (at zero applied flux). The resonance is fitted by the diameter-correction method~\cite{10.1063/1.4907935}, which yields the internal quality factor ${Q_i \approx 3800}$ and the coupling quality factor ${Q_c \approx 1000}$ when the ABSs are far detuned. We would like to emphasize that in our setup, the large coupling magnitude results in a significant magnetic flux dependence of the internal quality factor when the Andreev level energy approaches the bare resonator frequency. At ${\varphi = \pi}$ the nanowire introduces maximum losses to the system, and $Q_i$ usually drops to around 1000.

\section{Details of the fitting routine}

\subsection{SQPTs}

SQPT frequency can be found from Eq.~(\ref{long_dispersion}) (see the main text) as a difference between the energies of the two neighboring states. In our fitting routine, we first performed manual parameter selection to obtain a reasonable fit. Then, these parameters were used as an initial guess to the least-squares method. We found that the set of parameters is not always unique, which can produce large uncertainties. Furthermore, we frequently observed significant correlations between the parameters, such as the correlation between $x_0$ and $\Delta_{Al}$, which could reach almost 100\%. This is likely because both parameters are responsible for the overall energy scale. In order to mitigate these concerns, we fixed the gap parameter ${\Delta_{Al}/h = 52}$~GHz for all the fits depicted in Fig.~\ref{fig6}. This appeared to be a satisfactory initial parameter guess for each fit. 

\subsection{PTs}

The dispersion relation in Eq.~(\ref{long_dispersion}) can be simplified to describe PTs in a finite-length weak link. For a PT, we detect the average of a spin-split Andreev doublet and we can neglect spin-orbit coupling assuming that ${\Lambda_1 = \Lambda_2 = \Lambda}$. Expansion of Eq.~(\ref{long_dispersion}) around ${\varphi = \pi}$ up to second order in $\epsilon$ recovers the short-junction expression:

\begin{equation}
    E_A(\varphi) \approx \Delta'\sqrt{1-\tau \sin^2{(\varphi/2)}},
\end{equation}
with
\begin{equation}
    \Delta' = \frac{\Delta_{Al}}{\sqrt{(1+\Lambda)^2+(x_0 \Lambda \sqrt{1-\tau})^2}}.
\end{equation}

To fit the resonator frequency shift in single-tone spectra, we use the perturbative model developed in~\cite{PhysRevResearch.3.013036}. This model relies on expansion of the Hamiltonian up to second order in ${\Phi_{zpf}/\Phi_0}$, which is still applicable to our strongly coupled system. The resonator shift is determined by the ABSs in the ground state, which is described by
\begin{eqnarray}
    \delta f_{r} = -2\left(\frac{\pi\Phi_{zpf}}{\Phi_0}\right)^2 \pdv[2]{f_A}{\varphi}+\nonumber\\+\left(\frac{g_c(\varphi)}{2\pi}\right)^2\left(\frac{2}{f_A}-\frac{1}{f_A-f_r}-
    \frac{1}{f_A+f_r}\right),
    \label{eq1}
\end{eqnarray}  

with the phase-dependent coupling rate 
\begin{equation}
\frac{g_c(\varphi)}{2\pi} = \frac{\Phi_{zpf}}{\Phi_0} \sqrt{(1-\tau)} \left(-\pdv{f_A}{\varphi}\right) \tan(\varphi/2),
\label{eq2}
\end{equation}

where the pair transition frequency is expressed through its energy as ${f_A(\varphi) = 2E_A(\varphi)/h}$. Here, we introduced a single channel and neglected all the possible higher energy states, including the continuum of states above the superconducting gap, which is known to contribute to the inductive response of a finite-length weak link~
\cite{Kurilovich2021Nov, Fatemi2022Nov}. We also neglected the inductive energy of the shared inductance, which is generally true only when the supercurrent in the weak link is sufficiently small. However, this simple model demonstrated good correspondence between the extracted value of the coupling and the one directly observed in spectroscopy (e.g., in Fig.~\ref{fig4}, see the main text), which justifies its validity.

In our fits, we used the least squares method. When single-tone spectra in Fig.~\ref{fig4}(b) and Fig.~\ref{fig5}(b) were fitted, the data set was multiplied by ${f_A-f_r}$ to eliminate the divergence in Eq.~(\ref{eq1}). 

\section{Relation between the phase across the weak link and the applied flux}

In our devices, the loop inductance comprises the highly inductive NbTiN film with a total inductance of $2l\approx$~10~nH. When external flux is applied, part of the phase drops across the loop inductance, causing the self-screening effect~\cite{Jung2013Feb}:

\begin{equation}
    \varphi = \frac{2\pi}{\Phi_0} (\Phi - 2l I_s),
    \label{C1}
\end{equation}

where the supercurrent through the weak link is given by
\begin{equation}
    I_s = \frac{-2e}{\hbar}\sum_{p}\diffp{E^p_A}{\varphi}.
    \label{C2}
\end{equation}
Here, the sum is taken for all ABSs with dispersions $E^p_A$. The full spectrum of ABSs residing in the weak link spans up to $2 \Delta_{Al}$ and, therefore, is usually not accessible in experiments.

\begin{figure}
\includegraphics[width=\columnwidth]{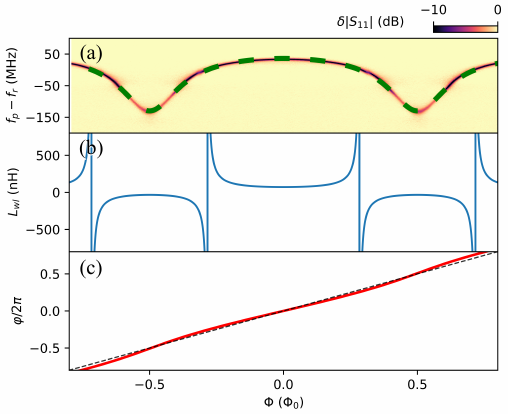}
\caption{(a) Single-tone spectrum measured in Device 3 at ${V_g = 1.5034}$~V offset to the bare resonator frequency ${f_r = 7.4685}$~GHz. The frequency shift fit as a function of applied flux is denoted by the green dashed line. (b) Corresponding weak link inductance, derived from (\ref{C3}, \ref{C4}). (c) The red line represents the phase across the weak link $\varphi$ as a function of the applied flux $\Phi$. For comparison, the black dashed line is plotted in the absence of self-screening ($2l$~=~0).}
\label{supC}
\end{figure}

However, it is possible to estimate the self-screening effect in the dispersive regime. Considering the circuit shown in Fig.~\ref{fig1}(d) classically, we obtain the resonator frequency shift that arises from shunting the $2l$ part of the resonator inductance with the phase-dependent weak link inductance $L_{wl}(\varphi)$:

\begin{equation}
    \frac{\delta f}{f_r} = -\frac{\delta L}{2 (2L_r)},
    \label{C3}
\end{equation}
where $2L_r$ is the total resonator inductance and 
\begin{equation}
    \delta L = 2L_r - 2l + \frac{2l L_{wl}}{2l + L_{wl}} - 2L_r = -\frac{(2l)^2}{2l+L_{wl}}.
    \label{C4}
\end{equation}
Thus, the weak link inductance can be directly retrieved from the single-tone spectra in dispersive regime. The total supercurrent is related to the weak link inductance as~\cite{PhysRevB.80.144520}:
\begin{equation}
    L_{wl}(\varphi)^{-1} = \frac{2\pi}{\Phi_0}\diffp{I_s}{\varphi}.
    \label{C5}
\end{equation}

We analyze the self-screening effect, using the single-tone spectroscopy data obtained for Device 3 at ${V_g = 1.5034}$~V, see Fig.~\ref{supC}(a). In this regime, no avoided crossing is observed, but the resonator shift is reasonably large (about 150~MHz), which implies that one or a few ABS with moderate transmission are present in the weak link, resulting in a relatively large supercurrent. The frequency shift fit is obtained by introducing one channel with ${\Delta'/h = 17}$~GHz, ${\tau = 0.68}$, ${\Phi_{zpf} = 0.025 \Phi_0}$. Using (\ref{C3}) and (\ref{C4}), we extract $L_{wl}$ as a function of the applied flux, see Fig.~\ref{supC}(b). Further, under the assumption of negligible self-screening ${\varphi \approx 2\pi \Phi/\Phi_0}$, we calculate the supercurrent by integrating (\ref{C5}), and plug it into (\ref{C1}) to estimate the self-screening. Fig.~\ref{supC}(c) shows that ${\varphi \approx 2\pi \Phi/\Phi_0}$ was indeed a reasonable approximation, and self-screening can be neglected.

\section{Power spectrum}

\begin{figure}[!htb]
\includegraphics[width=\columnwidth]{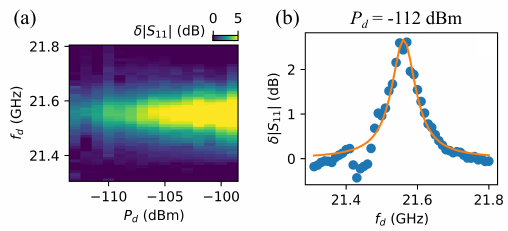}
\caption{(a) Power spectrum for the Device 3 at ${V_g = 3.962}$~V and ${\varphi = 0}$. (b) Line-cut of (a) at low power ${P_d = -112}$~dBm with a Lorentzian fit.}
\label{supE}
\end{figure}

Fig.~\ref{supE}(a) represents the power spectrum, obtained for one of the PT observed in Device 2 at $\varphi$~=~0. A low-power spectral line can be used to estimate the lower bound of the inhomogeneous dephasing time $T^*_2$ in this device~\cite{PhysRevLett.105.246804}. A line-cut at ${P_d = -112}$~dBm is shown in Fig.~\ref{supE}(b). A Lorentzian fit to the spectral line fit reveals the full width at half maximum ${f_\textrm{FWHM} \approx 82}$~MHz. We estimate the lower bound of the inhomogeneous dephasing time as ${T^*_2~\gtrapprox~1/\pi f_\textrm{FWHM} \approx 4}$~ns, a smaller value, compared to the previously reported~\cite{PhysRevLett.121.047001}. 

\section{Additional spectroscopy data}

\begin{figure}[!htb]
\includegraphics[width=\columnwidth]{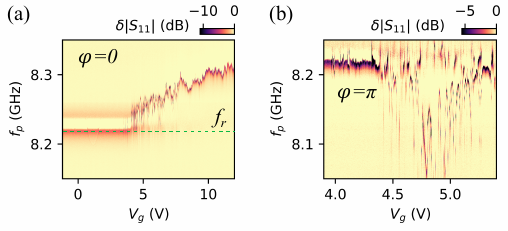}
\caption{Gate response measured in Device 1 at ${\varphi = 0}$ in (a) and at ${\varphi = \pi}$ in (b). The line at ${f_r=8.223}$~GHz in (a) denotes the bare resonator frequency, when the nanowire is fully pinched off. The gate-independent background was subtracted.}
\label{add_1}
\end{figure}

In this section, the supplementary data are provided.
\paragraph{Device 1.}
Fig.~\ref{add_1} shows the gate response of Device 1 at ${\varphi = 0}$ in (a) and at ${\varphi = \pi}$ in (b). 

\begin{figure}[!htb]
\includegraphics[width=\columnwidth]{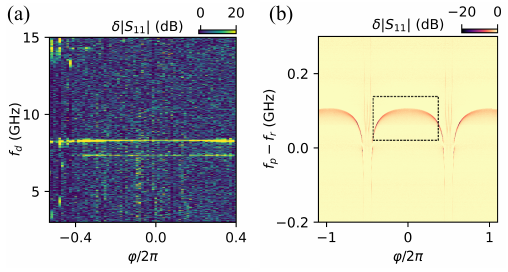}
\caption{Supplement to Fig.~\ref{fig5}. (a) Two-tone spectrum measured in Device 1 at ${V_g = 14.5}$~V. (b) Single-tone spectrum in a wide range of phase verifies the periodicity. The dashed rectangle denotes the range where (a) was measured.}
\label{add_2}
\end{figure}

Fig.~\ref{add_2} is a supplement Fig.~\ref{fig5}. The two-tone spectrum in (a) shows the lack of contributions from other higher-energy transitions, which justifies the use of the single-channel approximation in the fit in Fig.~\ref{fig5}(b). The wide-range single-tone spectrum in Fig.~\ref{add_2}(b) confirms the periodicity.

\begin{figure}[!htb]
\includegraphics[width=\columnwidth]{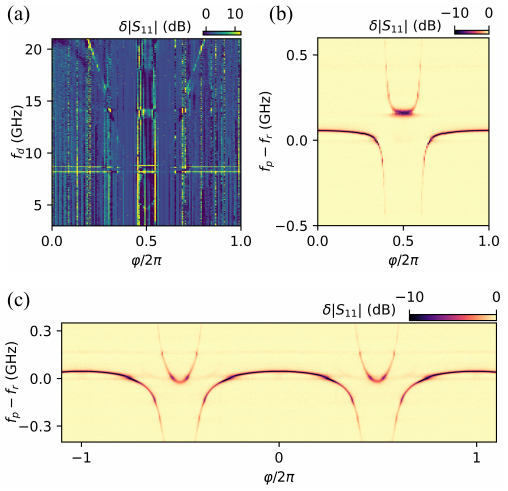}
\caption{Supplement to Fig.~\ref{fig4}. (a) Two-tone spectrum measured in Device 2 at ${V_g = 3.39}$~V. No higher energy transitions are observed, which validates the single-channel model used in the fits. (b) Corresponding single-tone spectrum. (c) Single-tone spectrum at ${V_g = 3.07}~V$ demonstrates periodicity.}
\label{add_3}
\end{figure}

\paragraph{Device 2.}
Fig.~\ref{add_3} is a supplement to Fig.~\ref{fig4}. The two-tone spectrum in (a) demonstrates the absence of higher-energy transitions, which justifies the single-channel approximation used in the fit in Fig.~\ref{fig4}(b). Panel (b) shows the corresponding single-tone spectrum. The wide range spectrum at ${V_g = 3.07}~V$  in (c) confirms the periodicity.

\begin{figure}[!htb]
\includegraphics[width=\columnwidth]{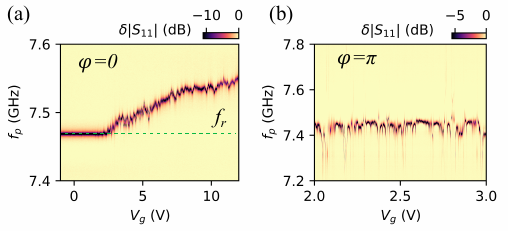}
\caption{Gate response measured in Device 3 at ${\varphi = 0}$ in (a) and at ${\varphi = \pi}$ in (b). The line at ${f_r=7.4685}$~GHz in (a) denotes the bare resonator frequency, when the nanowire is fully pinched off. The gate-independent background was subtracted.}
\label{add_4}
\end{figure}

\paragraph{Device 3.} Fig.~\ref{add_4} shows the gate response of Device 3 at ${\varphi = 0}$ in (a) and at ${\varphi = \pi}$ in (b).

\clearpage

\bibliography{libr}

\begin{thebibliography}{48}%
\makeatletter
\providecommand \@ifxundefined [1]{%
 \@ifx{#1\undefined}
}%
\providecommand \@ifnum [1]{%
 \ifnum #1\expandafter \@firstoftwo
 \else \expandafter \@secondoftwo
 \fi
}%
\providecommand \@ifx [1]{%
 \ifx #1\expandafter \@firstoftwo
 \else \expandafter \@secondoftwo
 \fi
}%
\providecommand \natexlab [1]{#1}%
\providecommand \enquote  [1]{``#1''}%
\providecommand \bibnamefont  [1]{#1}%
\providecommand \bibfnamefont [1]{#1}%
\providecommand \citenamefont [1]{#1}%
\providecommand \href@noop [0]{\@secondoftwo}%
\providecommand \href [0]{\begingroup \@sanitize@url \@href}%
\providecommand \@href[1]{\@@startlink{#1}\@@href}%
\providecommand \@@href[1]{\endgroup#1\@@endlink}%
\providecommand \@sanitize@url [0]{\catcode `\\12\catcode `\$12\catcode `\&12\catcode `\#12\catcode `\^12\catcode `\_12\catcode `\%12\relax}%
\providecommand \@@startlink[1]{}%
\providecommand \@@endlink[0]{}%
\providecommand \url  [0]{\begingroup\@sanitize@url \@url }%
\providecommand \@url [1]{\endgroup\@href {#1}{\urlprefix }}%
\providecommand \urlprefix  [0]{URL }%
\providecommand \Eprint [0]{\href }%
\providecommand \doibase [0]{https://doi.org/}%
\providecommand \selectlanguage [0]{\@gobble}%
\providecommand \bibinfo  [0]{\@secondoftwo}%
\providecommand \bibfield  [0]{\@secondoftwo}%
\providecommand \translation [1]{[#1]}%
\providecommand \BibitemOpen [0]{}%
\providecommand \bibitemStop [0]{}%
\providecommand \bibitemNoStop [0]{.\EOS\space}%
\providecommand \EOS [0]{\spacefactor3000\relax}%
\providecommand \BibitemShut  [1]{\csname bibitem#1\endcsname}%
\let\auto@bib@innerbib\@empty
%</preamble>
\bibitem [{\citenamefont {Blais}\ \emph {et~al.}(2021)\citenamefont {Blais}, \citenamefont {Grimsmo}, \citenamefont {Girvin},\ and\ \citenamefont {Wallraff}}]{RevModPhys.93.025005}%
  \BibitemOpen
  \bibfield  {author} {\bibinfo {author} {\bibfnamefont {A.}~\bibnamefont {Blais}}, \bibinfo {author} {\bibfnamefont {A.~L.}\ \bibnamefont {Grimsmo}}, \bibinfo {author} {\bibfnamefont {S.~M.}\ \bibnamefont {Girvin}},\ and\ \bibinfo {author} {\bibfnamefont {A.}~\bibnamefont {Wallraff}},\ }\bibfield  {title} {\bibinfo {title} {Circuit quantum electrodynamics},\ }\href {https://doi.org/10.1103/RevModPhys.93.025005} {\bibfield  {journal} {\bibinfo  {journal} {Rev. Mod. Phys.}\ }\textbf {\bibinfo {volume} {93}},\ \bibinfo {pages} {025005} (\bibinfo {year} {2021})}\BibitemShut {NoStop}%
\bibitem [{\citenamefont {Mi}\ \emph {et~al.}(2016)\citenamefont {Mi}, \citenamefont {Cady}, \citenamefont {Zajac}, \citenamefont {Deelman},\ and\ \citenamefont {Petta}}]{Mi2016Dec}%
  \BibitemOpen
  \bibfield  {author} {\bibinfo {author} {\bibfnamefont {X.}~\bibnamefont {Mi}}, \bibinfo {author} {\bibfnamefont {J.~V.}\ \bibnamefont {Cady}}, \bibinfo {author} {\bibfnamefont {D.~M.}\ \bibnamefont {Zajac}}, \bibinfo {author} {\bibfnamefont {P.~W.}\ \bibnamefont {Deelman}},\ and\ \bibinfo {author} {\bibfnamefont {J.~R.}\ \bibnamefont {Petta}},\ }\bibfield  {title} {\bibinfo {title} {{Strong coupling of a single electron in silicon to a microwave photon}},\ }\href {https://doi.org/10.1126/science.aal2469} {\bibfield  {journal} {\bibinfo  {journal} {Science}\ }\textbf {\bibinfo {volume} {355}},\ \bibinfo {pages} {156} (\bibinfo {year} {2016})}\BibitemShut {NoStop}%
\bibitem [{\citenamefont {{Clarke}}\ and\ \citenamefont {{Wilhelm}}(2008)}]{2008Natur.453.1031C}%
  \BibitemOpen
  \bibfield  {author} {\bibinfo {author} {\bibfnamefont {J.}~\bibnamefont {{Clarke}}}\ and\ \bibinfo {author} {\bibfnamefont {F.~K.}\ \bibnamefont {{Wilhelm}}},\ }\bibfield  {title} {\bibinfo {title} {{Superconducting quantum bits}},\ }\href {https://doi.org/10.1038/nature07128} {\bibfield  {journal} {\bibinfo  {journal} {Nature}\ }\textbf {\bibinfo {volume} {453}},\ \bibinfo {pages} {1031} (\bibinfo {year} {2008})}\BibitemShut {NoStop}%
\bibitem [{\citenamefont {Wallraff}\ \emph {et~al.}(2004)\citenamefont {Wallraff}, \citenamefont {Schuster}, \citenamefont {Blais}, \citenamefont {Frunzio}, \citenamefont {Huang}, \citenamefont {Majer}, \citenamefont {Kumar}, \citenamefont {Girvin},\ and\ \citenamefont {Schoelkopf}}]{Wallraff2004Sep}%
  \BibitemOpen
  \bibfield  {author} {\bibinfo {author} {\bibfnamefont {A.}~\bibnamefont {Wallraff}}, \bibinfo {author} {\bibfnamefont {D.~I.}\ \bibnamefont {Schuster}}, \bibinfo {author} {\bibfnamefont {A.}~\bibnamefont {Blais}}, \bibinfo {author} {\bibfnamefont {L.}~\bibnamefont {Frunzio}}, \bibinfo {author} {\bibfnamefont {R.-S.}\ \bibnamefont {Huang}}, \bibinfo {author} {\bibfnamefont {J.}~\bibnamefont {Majer}}, \bibinfo {author} {\bibfnamefont {S.}~\bibnamefont {Kumar}}, \bibinfo {author} {\bibfnamefont {S.~M.}\ \bibnamefont {Girvin}},\ and\ \bibinfo {author} {\bibfnamefont {R.~J.}\ \bibnamefont {Schoelkopf}},\ }\bibfield  {title} {\bibinfo {title} {{Strong coupling of a single photon to a superconducting qubit using circuit quantum electrodynamics}},\ }\href {https://doi.org/10.1038/nature02851} {\bibfield  {journal} {\bibinfo  {journal} {Nature}\ }\textbf {\bibinfo {volume} {431}},\ \bibinfo {pages} {162} (\bibinfo {year} {2004})}\BibitemShut {NoStop}%
\bibitem [{\citenamefont {Forn-D{\ifmmode\acute{\imath}\else\'{\i}\fi}az}\ \emph {et~al.}(2019)\citenamefont {Forn-D{\ifmmode\acute{\imath}\else\'{\i}\fi}az}, \citenamefont {Lamata}, \citenamefont {Rico}, \citenamefont {Kono},\ and\ \citenamefont {Solano}}]{Forn-Diaz2019Jun}%
  \BibitemOpen
  \bibfield  {author} {\bibinfo {author} {\bibfnamefont {P.}~\bibnamefont {Forn-D{\ifmmode\acute{\imath}\else\'{\i}\fi}az}}, \bibinfo {author} {\bibfnamefont {L.}~\bibnamefont {Lamata}}, \bibinfo {author} {\bibfnamefont {E.}~\bibnamefont {Rico}}, \bibinfo {author} {\bibfnamefont {J.}~\bibnamefont {Kono}},\ and\ \bibinfo {author} {\bibfnamefont {E.}~\bibnamefont {Solano}},\ }\bibfield  {title} {\bibinfo {title} {{Ultrastrong coupling regimes of light-matter interaction}},\ }\href {https://doi.org/10.1103/RevModPhys.91.025005} {\bibfield  {journal} {\bibinfo  {journal} {Rev. Mod. Phys.}\ }\textbf {\bibinfo {volume} {91}},\ \bibinfo {pages} {025005} (\bibinfo {year} {2019})}\BibitemShut {NoStop}%
\bibitem [{\citenamefont {Frisk~Kockum}\ \emph {et~al.}(2019)\citenamefont {Frisk~Kockum}, \citenamefont {Miranowicz}, \citenamefont {De~Liberato}, \citenamefont {Savasta},\ and\ \citenamefont {Nori}}]{FriskKockum2019Jan}%
  \BibitemOpen
  \bibfield  {author} {\bibinfo {author} {\bibfnamefont {A.}~\bibnamefont {Frisk~Kockum}}, \bibinfo {author} {\bibfnamefont {A.}~\bibnamefont {Miranowicz}}, \bibinfo {author} {\bibfnamefont {S.}~\bibnamefont {De~Liberato}}, \bibinfo {author} {\bibfnamefont {S.}~\bibnamefont {Savasta}},\ and\ \bibinfo {author} {\bibfnamefont {F.}~\bibnamefont {Nori}},\ }\bibfield  {title} {\bibinfo {title} {{Ultrastrong coupling between light and matter}},\ }\href {https://doi.org/10.1038/s42254-018-0006-2} {\bibfield  {journal} {\bibinfo  {journal} {Nat. Rev. Phys.}\ }\textbf {\bibinfo {volume} {1}},\ \bibinfo {pages} {19} (\bibinfo {year} {2019})}\BibitemShut {NoStop}%
\bibitem [{\citenamefont {Qin}\ \emph {et~al.}(2024)\citenamefont {Qin}, \citenamefont {Kockum}, \citenamefont {Mu{\ifmmode\tilde{n}\else\~{n}\fi}oz}, \citenamefont {Miranowicz},\ and\ \citenamefont {Nori}}]{Qin2024Aug}%
  \BibitemOpen
  \bibfield  {author} {\bibinfo {author} {\bibfnamefont {W.}~\bibnamefont {Qin}}, \bibinfo {author} {\bibfnamefont {A.~F.}\ \bibnamefont {Kockum}}, \bibinfo {author} {\bibfnamefont {C.~S.}\ \bibnamefont {Mu{\ifmmode\tilde{n}\else\~{n}\fi}oz}}, \bibinfo {author} {\bibfnamefont {A.}~\bibnamefont {Miranowicz}},\ and\ \bibinfo {author} {\bibfnamefont {F.}~\bibnamefont {Nori}},\ }\bibfield  {title} {\bibinfo {title} {{Quantum amplification and simulation of strong and ultrastrong coupling of light and matter}},\ }\href {https://doi.org/10.1016/j.physrep.2024.05.003} {\bibfield  {journal} {\bibinfo  {journal} {Phys. Rep.}\ }\textbf {\bibinfo {volume} {1078}},\ \bibinfo {pages} {1} (\bibinfo {year} {2024})}\BibitemShut {NoStop}%
\bibitem [{\citenamefont {Niemczyk}\ \emph {et~al.}(2010)\citenamefont {Niemczyk}, \citenamefont {Deppe}, \citenamefont {Huebl}, \citenamefont {Menzel}, \citenamefont {Hocke}, \citenamefont {Schwarz}, \citenamefont {Garcia-Ripoll}, \citenamefont {Zueco}, \citenamefont {H{\ifmmode\ddot{u}\else\"{u}\fi}mmer}, \citenamefont {Solano}, \citenamefont {Marx},\ and\ \citenamefont {Gross}}]{Niemczyk2010Oct}%
  \BibitemOpen
  \bibfield  {author} {\bibinfo {author} {\bibfnamefont {T.}~\bibnamefont {Niemczyk}}, \bibinfo {author} {\bibfnamefont {F.}~\bibnamefont {Deppe}}, \bibinfo {author} {\bibfnamefont {H.}~\bibnamefont {Huebl}}, \bibinfo {author} {\bibfnamefont {E.~P.}\ \bibnamefont {Menzel}}, \bibinfo {author} {\bibfnamefont {F.}~\bibnamefont {Hocke}}, \bibinfo {author} {\bibfnamefont {M.~J.}\ \bibnamefont {Schwarz}}, \bibinfo {author} {\bibfnamefont {J.~J.}\ \bibnamefont {Garcia-Ripoll}}, \bibinfo {author} {\bibfnamefont {D.}~\bibnamefont {Zueco}}, \bibinfo {author} {\bibfnamefont {T.}~\bibnamefont {H{\ifmmode\ddot{u}\else\"{u}\fi}mmer}}, \bibinfo {author} {\bibfnamefont {E.}~\bibnamefont {Solano}}, \bibinfo {author} {\bibfnamefont {A.}~\bibnamefont {Marx}},\ and\ \bibinfo {author} {\bibfnamefont {R.}~\bibnamefont {Gross}},\ }\bibfield  {title} {\bibinfo {title} {{Circuit quantum electrodynamics in the ultrastrong-coupling regime}},\ }\href {https://doi.org/10.1038/nphys1730} {\bibfield  {journal} {\bibinfo  {journal} {Nat.
  Phys.}\ }\textbf {\bibinfo {volume} {6}},\ \bibinfo {pages} {772} (\bibinfo {year} {2010})}\BibitemShut {NoStop}%
\bibitem [{\citenamefont {Wang}\ \emph {et~al.}(2017)\citenamefont {Wang}, \citenamefont {Guo}, \citenamefont {Zhang}, \citenamefont {Wang},\ and\ \citenamefont {Wu}}]{Wang2017Mar}%
  \BibitemOpen
  \bibfield  {author} {\bibinfo {author} {\bibfnamefont {Y.}~\bibnamefont {Wang}}, \bibinfo {author} {\bibfnamefont {C.}~\bibnamefont {Guo}}, \bibinfo {author} {\bibfnamefont {G.-Q.}\ \bibnamefont {Zhang}}, \bibinfo {author} {\bibfnamefont {G.}~\bibnamefont {Wang}},\ and\ \bibinfo {author} {\bibfnamefont {C.}~\bibnamefont {Wu}},\ }\bibfield  {title} {\bibinfo {title} {{Ultrafast quantum computation in ultrastrongly coupled circuit QED systems}},\ }\href {https://doi.org/10.1038/srep44251} {\bibfield  {journal} {\bibinfo  {journal} {Sci. Rep.}\ }\textbf {\bibinfo {volume} {7}},\ \bibinfo {pages} {1} (\bibinfo {year} {2017})}\BibitemShut {NoStop}%
\bibitem [{\citenamefont {{Kulik}}(1969)}]{1969JETP...30..944K}%
  \BibitemOpen
  \bibfield  {author} {\bibinfo {author} {\bibfnamefont {I.~O.}\ \bibnamefont {{Kulik}}},\ }\bibfield  {title} {\bibinfo {title} {{Macroscopic Quantization and the Proximity Effect in S-N-S Junctions}},\ }\href@noop {} {\bibfield  {journal} {\bibinfo  {journal} {Soviet Journal of Experimental and Theoretical Physics}\ }\textbf {\bibinfo {volume} {30}},\ \bibinfo {pages} {944} (\bibinfo {year} {1969})}\BibitemShut {NoStop}%
\bibitem [{\citenamefont {Beenakker}\ and\ \citenamefont {van Houten}(1991)}]{PhysRevLett.66.3056}%
  \BibitemOpen
  \bibfield  {author} {\bibinfo {author} {\bibfnamefont {C.~W.~J.}\ \bibnamefont {Beenakker}}\ and\ \bibinfo {author} {\bibfnamefont {H.}~\bibnamefont {van Houten}},\ }\bibfield  {title} {\bibinfo {title} {Josephson current through a superconducting quantum point contact shorter than the coherence length},\ }\href {https://doi.org/10.1103/PhysRevLett.66.3056} {\bibfield  {journal} {\bibinfo  {journal} {Phys. Rev. Lett.}\ }\textbf {\bibinfo {volume} {66}},\ \bibinfo {pages} {3056} (\bibinfo {year} {1991})}\BibitemShut {NoStop}%
\bibitem [{\citenamefont {Desp{\ifmmode\acute{o}\else\'{o}\fi}sito}\ and\ \citenamefont {Levy~Yeyati}(2001)}]{Desposito2001Sep}%
  \BibitemOpen
  \bibfield  {author} {\bibinfo {author} {\bibfnamefont {M.~A.}\ \bibnamefont {Desp{\ifmmode\acute{o}\else\'{o}\fi}sito}}\ and\ \bibinfo {author} {\bibfnamefont {A.}~\bibnamefont {Levy~Yeyati}},\ }\bibfield  {title} {\bibinfo {title} {{Controlled dephasing of Andreev states in superconducting quantum point contacts}},\ }\href {https://doi.org/10.1103/PhysRevB.64.140511} {\bibfield  {journal} {\bibinfo  {journal} {Phys. Rev. B}\ }\textbf {\bibinfo {volume} {64}},\ \bibinfo {pages} {140511} (\bibinfo {year} {2001})}\BibitemShut {NoStop}%
\bibitem [{\citenamefont {Zazunov}\ \emph {et~al.}(2003)\citenamefont {Zazunov}, \citenamefont {Shumeiko}, \citenamefont {Bratus'}, \citenamefont {Lantz},\ and\ \citenamefont {Wendin}}]{PhysRevLett.90.087003}%
  \BibitemOpen
  \bibfield  {author} {\bibinfo {author} {\bibfnamefont {A.}~\bibnamefont {Zazunov}}, \bibinfo {author} {\bibfnamefont {V.~S.}\ \bibnamefont {Shumeiko}}, \bibinfo {author} {\bibfnamefont {E.~N.}\ \bibnamefont {Bratus'}}, \bibinfo {author} {\bibfnamefont {J.}~\bibnamefont {Lantz}},\ and\ \bibinfo {author} {\bibfnamefont {G.}~\bibnamefont {Wendin}},\ }\bibfield  {title} {\bibinfo {title} {Andreev level qubit},\ }\href {https://doi.org/10.1103/PhysRevLett.90.087003} {\bibfield  {journal} {\bibinfo  {journal} {Phys. Rev. Lett.}\ }\textbf {\bibinfo {volume} {90}},\ \bibinfo {pages} {087003} (\bibinfo {year} {2003})}\BibitemShut {NoStop}%
\bibitem [{\citenamefont {Janvier}\ \emph {et~al.}(2015)\citenamefont {Janvier}, \citenamefont {Tosi}, \citenamefont {Bretheau}, \citenamefont {Girit}, \citenamefont {Stern}, \citenamefont {Bertet}, \citenamefont {Joyez}, \citenamefont {Vion}, \citenamefont {Esteve}, \citenamefont {Goffman}, \citenamefont {Pothier},\ and\ \citenamefont {Urbina}}]{Janvier2015Sep}%
  \BibitemOpen
  \bibfield  {author} {\bibinfo {author} {\bibfnamefont {C.}~\bibnamefont {Janvier}}, \bibinfo {author} {\bibfnamefont {L.}~\bibnamefont {Tosi}}, \bibinfo {author} {\bibfnamefont {L.}~\bibnamefont {Bretheau}}, \bibinfo {author} {\bibfnamefont {{\ifmmode\mbox{\c{C}}\else\c{C}\fi}.~{\ifmmode\ddot{O}\else\"{O}\fi}.}\ \bibnamefont {Girit}}, \bibinfo {author} {\bibfnamefont {M.}~\bibnamefont {Stern}}, \bibinfo {author} {\bibfnamefont {P.}~\bibnamefont {Bertet}}, \bibinfo {author} {\bibfnamefont {P.}~\bibnamefont {Joyez}}, \bibinfo {author} {\bibfnamefont {D.}~\bibnamefont {Vion}}, \bibinfo {author} {\bibfnamefont {D.}~\bibnamefont {Esteve}}, \bibinfo {author} {\bibfnamefont {M.~F.}\ \bibnamefont {Goffman}}, \bibinfo {author} {\bibfnamefont {H.}~\bibnamefont {Pothier}},\ and\ \bibinfo {author} {\bibfnamefont {C.}~\bibnamefont {Urbina}},\ }\bibfield  {title} {\bibinfo {title} {{Coherent manipulation of Andreev states in superconducting atomic contacts}},\ }\href {https://doi.org/10.1126/science.aab2179} {\bibfield
  {journal} {\bibinfo  {journal} {Science}\ }\textbf {\bibinfo {volume} {349}},\ \bibinfo {pages} {1199} (\bibinfo {year} {2015})}\BibitemShut {NoStop}%
\bibitem [{\citenamefont {Hays}\ \emph {et~al.}(2018)\citenamefont {Hays}, \citenamefont {de~Lange}, \citenamefont {Serniak}, \citenamefont {van Woerkom}, \citenamefont {Bouman}, \citenamefont {Krogstrup}, \citenamefont {Nyg\aa{}rd}, \citenamefont {Geresdi},\ and\ \citenamefont {Devoret}}]{PhysRevLett.121.047001}%
  \BibitemOpen
  \bibfield  {author} {\bibinfo {author} {\bibfnamefont {M.}~\bibnamefont {Hays}}, \bibinfo {author} {\bibfnamefont {G.}~\bibnamefont {de~Lange}}, \bibinfo {author} {\bibfnamefont {K.}~\bibnamefont {Serniak}}, \bibinfo {author} {\bibfnamefont {D.~J.}\ \bibnamefont {van Woerkom}}, \bibinfo {author} {\bibfnamefont {D.}~\bibnamefont {Bouman}}, \bibinfo {author} {\bibfnamefont {P.}~\bibnamefont {Krogstrup}}, \bibinfo {author} {\bibfnamefont {J.}~\bibnamefont {Nyg\aa{}rd}}, \bibinfo {author} {\bibfnamefont {A.}~\bibnamefont {Geresdi}},\ and\ \bibinfo {author} {\bibfnamefont {M.~H.}\ \bibnamefont {Devoret}},\ }\bibfield  {title} {\bibinfo {title} {Direct microwave measurement of andreev-bound-state dynamics in a semiconductor-nanowire josephson junction},\ }\href {https://doi.org/10.1103/PhysRevLett.121.047001} {\bibfield  {journal} {\bibinfo  {journal} {Phys. Rev. Lett.}\ }\textbf {\bibinfo {volume} {121}},\ \bibinfo {pages} {047001} (\bibinfo {year} {2018})}\BibitemShut {NoStop}%
\bibitem [{\citenamefont {Zellekens}\ \emph {et~al.}(2022)\citenamefont {Zellekens}, \citenamefont {Deacon}, \citenamefont {Perla}, \citenamefont {Gr{\ifmmode\ddot{u}\else\"{u}\fi}tzmacher}, \citenamefont {Lepsa}, \citenamefont {Sch{\ifmmode\ddot{a}\else\"{a}\fi}pers},\ and\ \citenamefont {Ishibashi}}]{Zellekens2022Oct}%
  \BibitemOpen
  \bibfield  {author} {\bibinfo {author} {\bibfnamefont {P.}~\bibnamefont {Zellekens}}, \bibinfo {author} {\bibfnamefont {R.~S.}\ \bibnamefont {Deacon}}, \bibinfo {author} {\bibfnamefont {P.}~\bibnamefont {Perla}}, \bibinfo {author} {\bibfnamefont {D.}~\bibnamefont {Gr{\ifmmode\ddot{u}\else\"{u}\fi}tzmacher}}, \bibinfo {author} {\bibfnamefont {M.~I.}\ \bibnamefont {Lepsa}}, \bibinfo {author} {\bibfnamefont {T.}~\bibnamefont {Sch{\ifmmode\ddot{a}\else\"{a}\fi}pers}},\ and\ \bibinfo {author} {\bibfnamefont {K.}~\bibnamefont {Ishibashi}},\ }\bibfield  {title} {\bibinfo {title} {{Microwave spectroscopy of Andreev states in InAs nanowire-based hybrid junctions using a flip-chip layout}},\ }\href {https://doi.org/10.1038/s42005-022-01035-6} {\bibfield  {journal} {\bibinfo  {journal} {Commun. Phys.}\ }\textbf {\bibinfo {volume} {5}},\ \bibinfo {pages} {1} (\bibinfo {year} {2022})}\BibitemShut {NoStop}%
\bibitem [{\citenamefont {Chtchelkatchev}\ and\ \citenamefont {Nazarov}(2003)}]{Chtchelkatchev2003Jun}%
  \BibitemOpen
  \bibfield  {author} {\bibinfo {author} {\bibfnamefont {N.~M.}\ \bibnamefont {Chtchelkatchev}}\ and\ \bibinfo {author} {\bibfnamefont {{\relax Yu}.~V.}\ \bibnamefont {Nazarov}},\ }\bibfield  {title} {\bibinfo {title} {{Andreev Quantum Dots for Spin Manipulation}},\ }\href {https://doi.org/10.1103/PhysRevLett.90.226806} {\bibfield  {journal} {\bibinfo  {journal} {Phys. Rev. Lett.}\ }\textbf {\bibinfo {volume} {90}},\ \bibinfo {pages} {226806} (\bibinfo {year} {2003})}\BibitemShut {NoStop}%
\bibitem [{\citenamefont {Padurariu}\ and\ \citenamefont {Nazarov}(2010)}]{Padurariu2010Apr}%
  \BibitemOpen
  \bibfield  {author} {\bibinfo {author} {\bibfnamefont {C.}~\bibnamefont {Padurariu}}\ and\ \bibinfo {author} {\bibfnamefont {{\relax Yu}.~V.}\ \bibnamefont {Nazarov}},\ }\bibfield  {title} {\bibinfo {title} {{Theoretical proposal for superconducting spin qubits}},\ }\href {https://doi.org/10.1103/PhysRevB.81.144519} {\bibfield  {journal} {\bibinfo  {journal} {Phys. Rev. B}\ }\textbf {\bibinfo {volume} {81}},\ \bibinfo {pages} {144519} (\bibinfo {year} {2010})}\BibitemShut {NoStop}%
\bibitem [{\citenamefont {Tosi}\ \emph {et~al.}(2019)\citenamefont {Tosi}, \citenamefont {Metzger}, \citenamefont {Goffman}, \citenamefont {Urbina}, \citenamefont {Pothier}, \citenamefont {Park}, \citenamefont {Yeyati}, \citenamefont {Nyg\aa{}rd},\ and\ \citenamefont {Krogstrup}}]{PhysRevX.9.011010}%
  \BibitemOpen
  \bibfield  {author} {\bibinfo {author} {\bibfnamefont {L.}~\bibnamefont {Tosi}}, \bibinfo {author} {\bibfnamefont {C.}~\bibnamefont {Metzger}}, \bibinfo {author} {\bibfnamefont {M.~F.}\ \bibnamefont {Goffman}}, \bibinfo {author} {\bibfnamefont {C.}~\bibnamefont {Urbina}}, \bibinfo {author} {\bibfnamefont {H.}~\bibnamefont {Pothier}}, \bibinfo {author} {\bibfnamefont {S.}~\bibnamefont {Park}}, \bibinfo {author} {\bibfnamefont {A.~L.}\ \bibnamefont {Yeyati}}, \bibinfo {author} {\bibfnamefont {J.}~\bibnamefont {Nyg\aa{}rd}},\ and\ \bibinfo {author} {\bibfnamefont {P.}~\bibnamefont {Krogstrup}},\ }\bibfield  {title} {\bibinfo {title} {Spin-orbit splitting of andreev states revealed by microwave spectroscopy},\ }\href {https://doi.org/10.1103/PhysRevX.9.011010} {\bibfield  {journal} {\bibinfo  {journal} {Phys. Rev. X}\ }\textbf {\bibinfo {volume} {9}},\ \bibinfo {pages} {011010} (\bibinfo {year} {2019})}\BibitemShut {NoStop}%
\bibitem [{\citenamefont {Hays}\ \emph {et~al.}(2021)\citenamefont {Hays}, \citenamefont {Fatemi}, \citenamefont {Bouman}, \citenamefont {Cerrillo}, \citenamefont {Diamond}, \citenamefont {Serniak}, \citenamefont {Connolly}, \citenamefont {Krogstrup}, \citenamefont {Nyg{\aa}rd}, \citenamefont {Yeyati}, \citenamefont {Geresdi},\ and\ \citenamefont {Devoret}}]{Hays2021Jul}%
  \BibitemOpen
  \bibfield  {author} {\bibinfo {author} {\bibfnamefont {M.}~\bibnamefont {Hays}}, \bibinfo {author} {\bibfnamefont {V.}~\bibnamefont {Fatemi}}, \bibinfo {author} {\bibfnamefont {D.}~\bibnamefont {Bouman}}, \bibinfo {author} {\bibfnamefont {J.}~\bibnamefont {Cerrillo}}, \bibinfo {author} {\bibfnamefont {S.}~\bibnamefont {Diamond}}, \bibinfo {author} {\bibfnamefont {K.}~\bibnamefont {Serniak}}, \bibinfo {author} {\bibfnamefont {T.}~\bibnamefont {Connolly}}, \bibinfo {author} {\bibfnamefont {P.}~\bibnamefont {Krogstrup}}, \bibinfo {author} {\bibfnamefont {J.}~\bibnamefont {Nyg{\aa}rd}}, \bibinfo {author} {\bibfnamefont {A.~L.}\ \bibnamefont {Yeyati}}, \bibinfo {author} {\bibfnamefont {A.}~\bibnamefont {Geresdi}},\ and\ \bibinfo {author} {\bibfnamefont {M.~H.}\ \bibnamefont {Devoret}},\ }\bibfield  {title} {\bibinfo {title} {{Coherent manipulation of an Andreev spin qubit}},\ }\href {https://doi.org/10.1126/science.abf0345} {\bibfield  {journal} {\bibinfo  {journal} {Science}\ }\textbf {\bibinfo {volume}
  {373}},\ \bibinfo {pages} {430} (\bibinfo {year} {2021})}\BibitemShut {NoStop}%
\bibitem [{\citenamefont {Hays}\ \emph {et~al.}(2020)\citenamefont {Hays}, \citenamefont {Fatemi}, \citenamefont {Serniak}, \citenamefont {Bouman}, \citenamefont {Diamond}, \citenamefont {de~Lange}, \citenamefont {Krogstrup}, \citenamefont {Nyg{\aa}rd}, \citenamefont {Geresdi},\ and\ \citenamefont {Devoret}}]{Hays2020Nov}%
  \BibitemOpen
  \bibfield  {author} {\bibinfo {author} {\bibfnamefont {M.}~\bibnamefont {Hays}}, \bibinfo {author} {\bibfnamefont {V.}~\bibnamefont {Fatemi}}, \bibinfo {author} {\bibfnamefont {K.}~\bibnamefont {Serniak}}, \bibinfo {author} {\bibfnamefont {D.}~\bibnamefont {Bouman}}, \bibinfo {author} {\bibfnamefont {S.}~\bibnamefont {Diamond}}, \bibinfo {author} {\bibfnamefont {G.}~\bibnamefont {de~Lange}}, \bibinfo {author} {\bibfnamefont {P.}~\bibnamefont {Krogstrup}}, \bibinfo {author} {\bibfnamefont {J.}~\bibnamefont {Nyg{\aa}rd}}, \bibinfo {author} {\bibfnamefont {A.}~\bibnamefont {Geresdi}},\ and\ \bibinfo {author} {\bibfnamefont {M.~H.}\ \bibnamefont {Devoret}},\ }\bibfield  {title} {\bibinfo {title} {{Continuous monitoring of a trapped superconducting spin}},\ }\href {https://doi.org/10.1038/s41567-020-0952-3} {\bibfield  {journal} {\bibinfo  {journal} {Nat. Phys.}\ }\textbf {\bibinfo {volume} {16}},\ \bibinfo {pages} {1103} (\bibinfo {year} {2020})}\BibitemShut {NoStop}%
\bibitem [{\citenamefont {Bargerbos}\ \emph {et~al.}(2023)\citenamefont {Bargerbos}, \citenamefont {Pita-Vidal}, \citenamefont {{\ifmmode\check{Z}\else\v{Z}\fi}itko}, \citenamefont {Splitthoff}, \citenamefont {Gr{\ifmmode\ddot{u}\else\"{u}\fi}nhaupt}, \citenamefont {Wesdorp}, \citenamefont {Liu}, \citenamefont {Kouwenhoven}, \citenamefont {Aguado}, \citenamefont {Andersen}, \citenamefont {Kou},\ and\ \citenamefont {van Heck}}]{Bargerbos2023Aug}%
  \BibitemOpen
  \bibfield  {author} {\bibinfo {author} {\bibfnamefont {A.}~\bibnamefont {Bargerbos}}, \bibinfo {author} {\bibfnamefont {M.}~\bibnamefont {Pita-Vidal}}, \bibinfo {author} {\bibfnamefont {R.}~\bibnamefont {{\ifmmode\check{Z}\else\v{Z}\fi}itko}}, \bibinfo {author} {\bibfnamefont {L.~J.}\ \bibnamefont {Splitthoff}}, \bibinfo {author} {\bibfnamefont {L.}~\bibnamefont {Gr{\ifmmode\ddot{u}\else\"{u}\fi}nhaupt}}, \bibinfo {author} {\bibfnamefont {J.~J.}\ \bibnamefont {Wesdorp}}, \bibinfo {author} {\bibfnamefont {Y.}~\bibnamefont {Liu}}, \bibinfo {author} {\bibfnamefont {L.~P.}\ \bibnamefont {Kouwenhoven}}, \bibinfo {author} {\bibfnamefont {R.}~\bibnamefont {Aguado}}, \bibinfo {author} {\bibfnamefont {C.~K.}\ \bibnamefont {Andersen}}, \bibinfo {author} {\bibfnamefont {A.}~\bibnamefont {Kou}},\ and\ \bibinfo {author} {\bibfnamefont {B.}~\bibnamefont {van Heck}},\ }\bibfield  {title} {\bibinfo {title} {{Spectroscopy of Spin-Split Andreev Levels in a Quantum Dot with Superconducting Leads}},\ }\href
  {https://doi.org/10.1103/PhysRevLett.131.097001} {\bibfield  {journal} {\bibinfo  {journal} {Phys. Rev. Lett.}\ }\textbf {\bibinfo {volume} {131}},\ \bibinfo {pages} {097001} (\bibinfo {year} {2023})}\BibitemShut {NoStop}%
\bibitem [{\citenamefont {Pita-Vidal}\ \emph {et~al.}(2023)\citenamefont {Pita-Vidal}, \citenamefont {Bargerbos}, \citenamefont {{\ifmmode\check{Z}\else\v{Z}\fi}itko}, \citenamefont {Splitthoff}, \citenamefont {Gr{\ifmmode\ddot{u}\else\"{u}\fi}nhaupt}, \citenamefont {Wesdorp}, \citenamefont {Liu}, \citenamefont {Kouwenhoven}, \citenamefont {Aguado}, \citenamefont {van Heck}, \citenamefont {Kou},\ and\ \citenamefont {Andersen}}]{Pita-Vidal2023Aug}%
  \BibitemOpen
  \bibfield  {author} {\bibinfo {author} {\bibfnamefont {M.}~\bibnamefont {Pita-Vidal}}, \bibinfo {author} {\bibfnamefont {A.}~\bibnamefont {Bargerbos}}, \bibinfo {author} {\bibfnamefont {R.}~\bibnamefont {{\ifmmode\check{Z}\else\v{Z}\fi}itko}}, \bibinfo {author} {\bibfnamefont {L.~J.}\ \bibnamefont {Splitthoff}}, \bibinfo {author} {\bibfnamefont {L.}~\bibnamefont {Gr{\ifmmode\ddot{u}\else\"{u}\fi}nhaupt}}, \bibinfo {author} {\bibfnamefont {J.~J.}\ \bibnamefont {Wesdorp}}, \bibinfo {author} {\bibfnamefont {Y.}~\bibnamefont {Liu}}, \bibinfo {author} {\bibfnamefont {L.~P.}\ \bibnamefont {Kouwenhoven}}, \bibinfo {author} {\bibfnamefont {R.}~\bibnamefont {Aguado}}, \bibinfo {author} {\bibfnamefont {B.}~\bibnamefont {van Heck}}, \bibinfo {author} {\bibfnamefont {A.}~\bibnamefont {Kou}},\ and\ \bibinfo {author} {\bibfnamefont {C.~K.}\ \bibnamefont {Andersen}},\ }\bibfield  {title} {\bibinfo {title} {{Direct manipulation of a superconducting spin qubit strongly coupled to a transmon qubit}},\ }\href
  {https://doi.org/10.1038/s41567-023-02071-x} {\bibfield  {journal} {\bibinfo  {journal} {Nat. Phys.}\ }\textbf {\bibinfo {volume} {19}},\ \bibinfo {pages} {1110} (\bibinfo {year} {2023})}\BibitemShut {NoStop}%
\bibitem [{\citenamefont {Metzger}\ \emph {et~al.}(2021)\citenamefont {Metzger}, \citenamefont {Park}, \citenamefont {Tosi}, \citenamefont {Janvier}, \citenamefont {Reynoso}, \citenamefont {Goffman}, \citenamefont {Urbina}, \citenamefont {Levy~Yeyati},\ and\ \citenamefont {Pothier}}]{PhysRevResearch.3.013036}%
  \BibitemOpen
  \bibfield  {author} {\bibinfo {author} {\bibfnamefont {C.}~\bibnamefont {Metzger}}, \bibinfo {author} {\bibfnamefont {S.}~\bibnamefont {Park}}, \bibinfo {author} {\bibfnamefont {L.}~\bibnamefont {Tosi}}, \bibinfo {author} {\bibfnamefont {C.}~\bibnamefont {Janvier}}, \bibinfo {author} {\bibfnamefont {A.~A.}\ \bibnamefont {Reynoso}}, \bibinfo {author} {\bibfnamefont {M.~F.}\ \bibnamefont {Goffman}}, \bibinfo {author} {\bibfnamefont {C.}~\bibnamefont {Urbina}}, \bibinfo {author} {\bibfnamefont {A.}~\bibnamefont {Levy~Yeyati}},\ and\ \bibinfo {author} {\bibfnamefont {H.}~\bibnamefont {Pothier}},\ }\bibfield  {title} {\bibinfo {title} {{Circuit-QED with phase-biased Josephson weak links}},\ }\href {https://doi.org/10.1103/PhysRevResearch.3.013036} {\bibfield  {journal} {\bibinfo  {journal} {Phys. Rev. Res.}\ }\textbf {\bibinfo {volume} {3}},\ \bibinfo {pages} {013036} (\bibinfo {year} {2021})}\BibitemShut {NoStop}%
\bibitem [{\citenamefont {Matute-Ca{\ifmmode\tilde{n}\else\~{n}\fi}adas}\ \emph {et~al.}(2022)\citenamefont {Matute-Ca{\ifmmode\tilde{n}\else\~{n}\fi}adas}, \citenamefont {Metzger}, \citenamefont {Park}, \citenamefont {Tosi}, \citenamefont {Krogstrup}, \citenamefont {Nyg{\aa}rd}, \citenamefont {Goffman}, \citenamefont {Urbina}, \citenamefont {Pothier},\ and\ \citenamefont {Yeyati}}]{Matute-Canadas2022May}%
  \BibitemOpen
  \bibfield  {author} {\bibinfo {author} {\bibfnamefont {F.~J.}\ \bibnamefont {Matute-Ca{\ifmmode\tilde{n}\else\~{n}\fi}adas}}, \bibinfo {author} {\bibfnamefont {C.}~\bibnamefont {Metzger}}, \bibinfo {author} {\bibfnamefont {S.}~\bibnamefont {Park}}, \bibinfo {author} {\bibfnamefont {L.}~\bibnamefont {Tosi}}, \bibinfo {author} {\bibfnamefont {P.}~\bibnamefont {Krogstrup}}, \bibinfo {author} {\bibfnamefont {J.}~\bibnamefont {Nyg{\aa}rd}}, \bibinfo {author} {\bibfnamefont {M.~F.}\ \bibnamefont {Goffman}}, \bibinfo {author} {\bibfnamefont {C.}~\bibnamefont {Urbina}}, \bibinfo {author} {\bibfnamefont {H.}~\bibnamefont {Pothier}},\ and\ \bibinfo {author} {\bibfnamefont {A.~L.}\ \bibnamefont {Yeyati}},\ }\bibfield  {title} {\bibinfo {title} {{Signatures of Interactions in the Andreev Spectrum of Nanowire Josephson Junctions}},\ }\href {https://doi.org/10.1103/PhysRevLett.128.197702} {\bibfield  {journal} {\bibinfo  {journal} {Phys. Rev. Lett.}\ }\textbf {\bibinfo {volume} {128}},\ \bibinfo {pages} {197702}
  (\bibinfo {year} {2022})}\BibitemShut {NoStop}%
\bibitem [{\citenamefont {Wesdorp}\ \emph {et~al.}(2024)\citenamefont {Wesdorp}, \citenamefont {Matute-Ca{\ifmmode\tilde{n}\else\~{n}\fi}adas}, \citenamefont {Vaartjes}, \citenamefont {Gr{\ifmmode\ddot{u}\else\"{u}\fi}nhaupt}, \citenamefont {Laeven}, \citenamefont {Roelofs}, \citenamefont {Splitthoff}, \citenamefont {Pita-Vidal}, \citenamefont {Bargerbos}, \citenamefont {van Woerkom}, \citenamefont {Krogstrup}, \citenamefont {Kouwenhoven}, \citenamefont {Andersen}, \citenamefont {Yeyati}, \citenamefont {van Heck},\ and\ \citenamefont {de~Lange}}]{Wesdorp2024Jan}%
  \BibitemOpen
  \bibfield  {author} {\bibinfo {author} {\bibfnamefont {J.~J.}\ \bibnamefont {Wesdorp}}, \bibinfo {author} {\bibfnamefont {F.~J.}\ \bibnamefont {Matute-Ca{\ifmmode\tilde{n}\else\~{n}\fi}adas}}, \bibinfo {author} {\bibfnamefont {A.}~\bibnamefont {Vaartjes}}, \bibinfo {author} {\bibfnamefont {L.}~\bibnamefont {Gr{\ifmmode\ddot{u}\else\"{u}\fi}nhaupt}}, \bibinfo {author} {\bibfnamefont {T.}~\bibnamefont {Laeven}}, \bibinfo {author} {\bibfnamefont {S.}~\bibnamefont {Roelofs}}, \bibinfo {author} {\bibfnamefont {L.~J.}\ \bibnamefont {Splitthoff}}, \bibinfo {author} {\bibfnamefont {M.}~\bibnamefont {Pita-Vidal}}, \bibinfo {author} {\bibfnamefont {A.}~\bibnamefont {Bargerbos}}, \bibinfo {author} {\bibfnamefont {D.~J.}\ \bibnamefont {van Woerkom}}, \bibinfo {author} {\bibfnamefont {P.}~\bibnamefont {Krogstrup}}, \bibinfo {author} {\bibfnamefont {L.~P.}\ \bibnamefont {Kouwenhoven}}, \bibinfo {author} {\bibfnamefont {C.~K.}\ \bibnamefont {Andersen}}, \bibinfo {author} {\bibfnamefont {A.~L.}\ \bibnamefont {Yeyati}},
  \bibinfo {author} {\bibfnamefont {B.}~\bibnamefont {van Heck}},\ and\ \bibinfo {author} {\bibfnamefont {G.}~\bibnamefont {de~Lange}},\ }\bibfield  {title} {\bibinfo {title} {{Microwave spectroscopy of interacting Andreev spins}},\ }\href {https://doi.org/10.1103/PhysRevB.109.045302} {\bibfield  {journal} {\bibinfo  {journal} {Phys. Rev. B}\ }\textbf {\bibinfo {volume} {109}},\ \bibinfo {pages} {045302} (\bibinfo {year} {2024})}\BibitemShut {NoStop}%
\bibitem [{\citenamefont {Krogstrup}\ \emph {et~al.}(2015)\citenamefont {Krogstrup}, \citenamefont {Ziino}, \citenamefont {Chang}, \citenamefont {Albrecht}, \citenamefont {Madsen}, \citenamefont {Johnson}, \citenamefont {Nyg{\aa}rd}, \citenamefont {Marcus},\ and\ \citenamefont {Jespersen}}]{Krogstrup2015Apr}%
  \BibitemOpen
  \bibfield  {author} {\bibinfo {author} {\bibfnamefont {P.}~\bibnamefont {Krogstrup}}, \bibinfo {author} {\bibfnamefont {N.~L.~B.}\ \bibnamefont {Ziino}}, \bibinfo {author} {\bibfnamefont {W.}~\bibnamefont {Chang}}, \bibinfo {author} {\bibfnamefont {S.~M.}\ \bibnamefont {Albrecht}}, \bibinfo {author} {\bibfnamefont {M.~H.}\ \bibnamefont {Madsen}}, \bibinfo {author} {\bibfnamefont {E.}~\bibnamefont {Johnson}}, \bibinfo {author} {\bibfnamefont {J.}~\bibnamefont {Nyg{\aa}rd}}, \bibinfo {author} {\bibfnamefont {C.~M.}\ \bibnamefont {Marcus}},\ and\ \bibinfo {author} {\bibfnamefont {T.~S.}\ \bibnamefont {Jespersen}},\ }\bibfield  {title} {\bibinfo {title} {{Epitaxy of semiconductor{\textendash}superconductor nanowires}},\ }\href {https://doi.org/10.1038/nmat4176} {\bibfield  {journal} {\bibinfo  {journal} {Nat. Mater.}\ }\textbf {\bibinfo {volume} {14}},\ \bibinfo {pages} {400} (\bibinfo {year} {2015})}\BibitemShut {NoStop}%
\bibitem [{\citenamefont {Pita-Vidal}\ \emph {et~al.}(2025)\citenamefont {Pita-Vidal}, \citenamefont {Wesdorp},\ and\ \citenamefont {Andersen}}]{Pita-Vidal2025Jan}%
  \BibitemOpen
  \bibfield  {author} {\bibinfo {author} {\bibfnamefont {M.}~\bibnamefont {Pita-Vidal}}, \bibinfo {author} {\bibfnamefont {J.~J.}\ \bibnamefont {Wesdorp}},\ and\ \bibinfo {author} {\bibfnamefont {C.~K.}\ \bibnamefont {Andersen}},\ }\bibfield  {title} {\bibinfo {title} {{Blueprint for All-to-All-Connected Superconducting Spin Qubits}},\ }\href {https://doi.org/10.1103/PRXQuantum.6.010308} {\bibfield  {journal} {\bibinfo  {journal} {PRX Quantum}\ }\textbf {\bibinfo {volume} {6}},\ \bibinfo {pages} {010308} (\bibinfo {year} {2025})}\BibitemShut {NoStop}%
\bibitem [{\citenamefont {Lu}\ \emph {et~al.}(2024)\citenamefont {Lu}, \citenamefont {Day}, \citenamefont {Akhmerov}, \citenamefont {van Heck},\ and\ \citenamefont {Fatemi}}]{Lu2024Dec}%
  \BibitemOpen
  \bibfield  {author} {\bibinfo {author} {\bibfnamefont {H.}~\bibnamefont {Lu}}, \bibinfo {author} {\bibfnamefont {I.~A.}\ \bibnamefont {Day}}, \bibinfo {author} {\bibfnamefont {A.~R.}\ \bibnamefont {Akhmerov}}, \bibinfo {author} {\bibfnamefont {B.}~\bibnamefont {van Heck}},\ and\ \bibinfo {author} {\bibfnamefont {V.}~\bibnamefont {Fatemi}},\ }\bibfield  {title} {\bibinfo {title} {{Kramers-protected hardware-efficient error correction with Andreev spin qubits}},\ }\bibfield  {journal} {\bibinfo  {journal} {arXiv}\ }\href {https://doi.org/10.48550/arXiv.2412.16116} {10.48550/arXiv.2412.16116} (\bibinfo {year} {2024}),\ \Eprint {https://arxiv.org/abs/2412.16116} {2412.16116} \BibitemShut {NoStop}%
\bibitem [{\citenamefont {Abay}\ \emph {et~al.}(2014)\citenamefont {Abay}, \citenamefont {Persson}, \citenamefont {Nilsson}, \citenamefont {Wu}, \citenamefont {Xu}, \citenamefont {Fogelstr\"om}, \citenamefont {Shumeiko},\ and\ \citenamefont {Delsing}}]{PhysRevB.89.214508}%
  \BibitemOpen
  \bibfield  {author} {\bibinfo {author} {\bibfnamefont {S.}~\bibnamefont {Abay}}, \bibinfo {author} {\bibfnamefont {D.}~\bibnamefont {Persson}}, \bibinfo {author} {\bibfnamefont {H.}~\bibnamefont {Nilsson}}, \bibinfo {author} {\bibfnamefont {F.}~\bibnamefont {Wu}}, \bibinfo {author} {\bibfnamefont {H.~Q.}\ \bibnamefont {Xu}}, \bibinfo {author} {\bibfnamefont {M.}~\bibnamefont {Fogelstr\"om}}, \bibinfo {author} {\bibfnamefont {V.}~\bibnamefont {Shumeiko}},\ and\ \bibinfo {author} {\bibfnamefont {P.}~\bibnamefont {Delsing}},\ }\bibfield  {title} {\bibinfo {title} {Charge transport in inas nanowire josephson junctions},\ }\href {https://doi.org/10.1103/PhysRevB.89.214508} {\bibfield  {journal} {\bibinfo  {journal} {Phys. Rev. B}\ }\textbf {\bibinfo {volume} {89}},\ \bibinfo {pages} {214508} (\bibinfo {year} {2014})}\BibitemShut {NoStop}%
\bibitem [{\citenamefont {Devoret}\ \emph {et~al.}(2007)\citenamefont {Devoret}, \citenamefont {Girvin},\ and\ \citenamefont {Schoelkopf}}]{Devoret2007Oct}%
  \BibitemOpen
  \bibfield  {author} {\bibinfo {author} {\bibfnamefont {M.~H.}\ \bibnamefont {Devoret}}, \bibinfo {author} {\bibfnamefont {S.}~\bibnamefont {Girvin}},\ and\ \bibinfo {author} {\bibfnamefont {R.}~\bibnamefont {Schoelkopf}},\ }\bibfield  {title} {\bibinfo {title} {{Circuit-QED: How strong can the coupling between a Josephson junction atom and a transmission line resonator be?}},\ }\href {https://doi.org/10.1002/andp.200751910-1109} {\bibfield  {journal} {\bibinfo  {journal} {Ann. Phys.}\ }\textbf {\bibinfo {volume} {519}},\ \bibinfo {pages} {767} (\bibinfo {year} {2007})}\BibitemShut {NoStop}%
\bibitem [{\citenamefont {Stockklauser}\ \emph {et~al.}(2017)\citenamefont {Stockklauser}, \citenamefont {Scarlino}, \citenamefont {Koski}, \citenamefont {Gasparinetti}, \citenamefont {Andersen}, \citenamefont {Reichl}, \citenamefont {Wegscheider}, \citenamefont {Ihn}, \citenamefont {Ensslin},\ and\ \citenamefont {Wallraff}}]{Stockklauser2017Mar}%
  \BibitemOpen
  \bibfield  {author} {\bibinfo {author} {\bibfnamefont {A.}~\bibnamefont {Stockklauser}}, \bibinfo {author} {\bibfnamefont {P.}~\bibnamefont {Scarlino}}, \bibinfo {author} {\bibfnamefont {J.~V.}\ \bibnamefont {Koski}}, \bibinfo {author} {\bibfnamefont {S.}~\bibnamefont {Gasparinetti}}, \bibinfo {author} {\bibfnamefont {C.~K.}\ \bibnamefont {Andersen}}, \bibinfo {author} {\bibfnamefont {C.}~\bibnamefont {Reichl}}, \bibinfo {author} {\bibfnamefont {W.}~\bibnamefont {Wegscheider}}, \bibinfo {author} {\bibfnamefont {T.}~\bibnamefont {Ihn}}, \bibinfo {author} {\bibfnamefont {K.}~\bibnamefont {Ensslin}},\ and\ \bibinfo {author} {\bibfnamefont {A.}~\bibnamefont {Wallraff}},\ }\bibfield  {title} {\bibinfo {title} {{Strong Coupling Cavity QED with Gate-Defined Double Quantum Dots Enabled by a High Impedance Resonator}},\ }\href {https://doi.org/10.1103/PhysRevX.7.011030} {\bibfield  {journal} {\bibinfo  {journal} {Phys. Rev. X}\ }\textbf {\bibinfo {volume} {7}},\ \bibinfo {pages} {011030} (\bibinfo {year}
  {2017})}\BibitemShut {NoStop}%
\bibitem [{\citenamefont {Sestoft}\ \emph {et~al.}(2024)\citenamefont {Sestoft}, \citenamefont {Marnauza}, \citenamefont {Olsteins}, \citenamefont {Kanne}, \citenamefont {Schlosser}, \citenamefont {Chen}, \citenamefont {Grove-rasmussen},\ and\ \citenamefont {Nyg{\aa}rd}}]{Sestoft2024Jun}%
  \BibitemOpen
  \bibfield  {author} {\bibinfo {author} {\bibfnamefont {J.~E.}\ \bibnamefont {Sestoft}}, \bibinfo {author} {\bibfnamefont {M.}~\bibnamefont {Marnauza}}, \bibinfo {author} {\bibfnamefont {D.}~\bibnamefont {Olsteins}}, \bibinfo {author} {\bibfnamefont {T.}~\bibnamefont {Kanne}}, \bibinfo {author} {\bibfnamefont {R.~D.}\ \bibnamefont {Schlosser}}, \bibinfo {author} {\bibfnamefont {I.-j.}\ \bibnamefont {Chen}}, \bibinfo {author} {\bibfnamefont {K.}~\bibnamefont {Grove-rasmussen}},\ and\ \bibinfo {author} {\bibfnamefont {J.}~\bibnamefont {Nyg{\aa}rd}},\ }\bibfield  {title} {\bibinfo {title} {{Shadowed versus Etched Superconductor{\textendash}Semiconductor Junctions in Al/InAs Nanowires}},\ }\href {https://doi.org/10.1021/acs.nanolett.4c02055} {\bibfield  {journal} {\bibinfo  {journal} {Nano Lett.}\ }\textbf {\bibinfo {volume} {24}},\ \bibinfo {pages} {8394} (\bibinfo {year} {2024})}\BibitemShut {NoStop}%
\bibitem [{\citenamefont {Park}\ and\ \citenamefont {Yeyati}(2017)}]{Park2017Sep}%
  \BibitemOpen
  \bibfield  {author} {\bibinfo {author} {\bibfnamefont {S.}~\bibnamefont {Park}}\ and\ \bibinfo {author} {\bibfnamefont {A.~L.}\ \bibnamefont {Yeyati}},\ }\bibfield  {title} {\bibinfo {title} {{Andreev spin qubits in multichannel Rashba nanowires}},\ }\href {https://doi.org/10.1103/PhysRevB.96.125416} {\bibfield  {journal} {\bibinfo  {journal} {Phys. Rev. B}\ }\textbf {\bibinfo {volume} {96}},\ \bibinfo {pages} {125416} (\bibinfo {year} {2017})}\BibitemShut {NoStop}%
\bibitem [{\citenamefont {Ungerer}\ \emph {et~al.}(2024)\citenamefont {Ungerer}, \citenamefont {Pally}, \citenamefont {Kononov}, \citenamefont {Lehmann}, \citenamefont {Ridderbos}, \citenamefont {Potts}, \citenamefont {Thelander}, \citenamefont {Dick}, \citenamefont {Maisi}, \citenamefont {Scarlino}, \citenamefont {Baumgartner},\ and\ \citenamefont {Sch{\ifmmode\ddot{o}\else\"{o}\fi}nenberger}}]{Ungerer2024Feb}%
  \BibitemOpen
  \bibfield  {author} {\bibinfo {author} {\bibfnamefont {J.~H.}\ \bibnamefont {Ungerer}}, \bibinfo {author} {\bibfnamefont {A.}~\bibnamefont {Pally}}, \bibinfo {author} {\bibfnamefont {A.}~\bibnamefont {Kononov}}, \bibinfo {author} {\bibfnamefont {S.}~\bibnamefont {Lehmann}}, \bibinfo {author} {\bibfnamefont {J.}~\bibnamefont {Ridderbos}}, \bibinfo {author} {\bibfnamefont {P.~P.}\ \bibnamefont {Potts}}, \bibinfo {author} {\bibfnamefont {C.}~\bibnamefont {Thelander}}, \bibinfo {author} {\bibfnamefont {K.~A.}\ \bibnamefont {Dick}}, \bibinfo {author} {\bibfnamefont {V.~F.}\ \bibnamefont {Maisi}}, \bibinfo {author} {\bibfnamefont {P.}~\bibnamefont {Scarlino}}, \bibinfo {author} {\bibfnamefont {A.}~\bibnamefont {Baumgartner}},\ and\ \bibinfo {author} {\bibfnamefont {C.}~\bibnamefont {Sch{\ifmmode\ddot{o}\else\"{o}\fi}nenberger}},\ }\bibfield  {title} {\bibinfo {title} {{Strong coupling between a microwave photon and a singlet-triplet qubit}},\ }\href {https://doi.org/10.1038/s41467-024-45235-w} {\bibfield
  {journal} {\bibinfo  {journal} {Nat. Commun.}\ }\textbf {\bibinfo {volume} {15}},\ \bibinfo {pages} {1} (\bibinfo {year} {2024})}\BibitemShut {NoStop}%
\bibitem [{\citenamefont {Fatemi}\ \emph {et~al.}(2022)\citenamefont {Fatemi}, \citenamefont {Kurilovich}, \citenamefont {Hays}, \citenamefont {Bouman}, \citenamefont {Connolly}, \citenamefont {Diamond}, \citenamefont {Frattini}, \citenamefont {Kurilovich}, \citenamefont {Krogstrup}, \citenamefont {Nyg{\aa}rd}, \citenamefont {Geresdi}, \citenamefont {Glazman},\ and\ \citenamefont {Devoret}}]{Fatemi2022Nov}%
  \BibitemOpen
  \bibfield  {author} {\bibinfo {author} {\bibfnamefont {V.}~\bibnamefont {Fatemi}}, \bibinfo {author} {\bibfnamefont {P.~D.}\ \bibnamefont {Kurilovich}}, \bibinfo {author} {\bibfnamefont {M.}~\bibnamefont {Hays}}, \bibinfo {author} {\bibfnamefont {D.}~\bibnamefont {Bouman}}, \bibinfo {author} {\bibfnamefont {T.}~\bibnamefont {Connolly}}, \bibinfo {author} {\bibfnamefont {S.}~\bibnamefont {Diamond}}, \bibinfo {author} {\bibfnamefont {N.~E.}\ \bibnamefont {Frattini}}, \bibinfo {author} {\bibfnamefont {V.~D.}\ \bibnamefont {Kurilovich}}, \bibinfo {author} {\bibfnamefont {P.}~\bibnamefont {Krogstrup}}, \bibinfo {author} {\bibfnamefont {J.}~\bibnamefont {Nyg{\aa}rd}}, \bibinfo {author} {\bibfnamefont {A.}~\bibnamefont {Geresdi}}, \bibinfo {author} {\bibfnamefont {L.~I.}\ \bibnamefont {Glazman}},\ and\ \bibinfo {author} {\bibfnamefont {M.~H.}\ \bibnamefont {Devoret}},\ }\bibfield  {title} {\bibinfo {title} {{Microwave Susceptibility Observation of Interacting Many-Body Andreev States}},\ }\href
  {https://doi.org/10.1103/PhysRevLett.129.227701} {\bibfield  {journal} {\bibinfo  {journal} {Phys. Rev. Lett.}\ }\textbf {\bibinfo {volume} {129}},\ \bibinfo {pages} {227701} (\bibinfo {year} {2022})}\BibitemShut {NoStop}%
\bibitem [{\citenamefont {Sahu}\ \emph {et~al.}(2024)\citenamefont {Sahu}, \citenamefont {Matute-Ca{\ifmmode\tilde{n}\else\~{n}\fi}adas}, \citenamefont {Benito}, \citenamefont {Krogstrup}, \citenamefont {Nyg{\aa}rd}, \citenamefont {Goffman}, \citenamefont {Urbina}, \citenamefont {Yeyati},\ and\ \citenamefont {Pothier}}]{Sahu2024Apr}%
  \BibitemOpen
  \bibfield  {author} {\bibinfo {author} {\bibfnamefont {M.~R.}\ \bibnamefont {Sahu}}, \bibinfo {author} {\bibfnamefont {F.~J.}\ \bibnamefont {Matute-Ca{\ifmmode\tilde{n}\else\~{n}\fi}adas}}, \bibinfo {author} {\bibfnamefont {M.}~\bibnamefont {Benito}}, \bibinfo {author} {\bibfnamefont {P.}~\bibnamefont {Krogstrup}}, \bibinfo {author} {\bibfnamefont {J.}~\bibnamefont {Nyg{\aa}rd}}, \bibinfo {author} {\bibfnamefont {M.~F.}\ \bibnamefont {Goffman}}, \bibinfo {author} {\bibfnamefont {C.}~\bibnamefont {Urbina}}, \bibinfo {author} {\bibfnamefont {A.~L.}\ \bibnamefont {Yeyati}},\ and\ \bibinfo {author} {\bibfnamefont {H.}~\bibnamefont {Pothier}},\ }\bibfield  {title} {\bibinfo {title} {{Ground-state phase diagram and parity-flipping microwave transitions in a gate-tunable Josephson junction}},\ }\href {https://doi.org/10.1103/PhysRevB.109.134506} {\bibfield  {journal} {\bibinfo  {journal} {Phys. Rev. B}\ }\textbf {\bibinfo {volume} {109}},\ \bibinfo {pages} {134506} (\bibinfo {year} {2024})}\BibitemShut {NoStop}%
\bibitem [{\citenamefont {METZGER}(2022)}]{METZGER:2022iuj}%
  \BibitemOpen
  \bibfield  {author} {\bibinfo {author} {\bibfnamefont {C.}~\bibnamefont {METZGER}},\ }\emph {\bibinfo {title} {{Spin \& charge effects in Andreev Bound States}}},\ \href@noop {} {Ph.D. thesis},\ \bibinfo  {school} {Service de physique de l'\'etat condens\'e, France, Quantronics Group, France, Saclay} (\bibinfo {year} {2022})\BibitemShut {NoStop}%
\bibitem [{\citenamefont {Lu}\ \emph {et~al.}(2025)\citenamefont {Lu}, \citenamefont {Bofill}, \citenamefont {Sun}, \citenamefont {Kanne}, \citenamefont {Nyg{\aa}rd}, \citenamefont {Kjaergaard},\ and\ \citenamefont {Fatemi}}]{Lu2025Jan}%
  \BibitemOpen
  \bibfield  {author} {\bibinfo {author} {\bibfnamefont {H.}~\bibnamefont {Lu}}, \bibinfo {author} {\bibfnamefont {D.~F.}\ \bibnamefont {Bofill}}, \bibinfo {author} {\bibfnamefont {Z.}~\bibnamefont {Sun}}, \bibinfo {author} {\bibfnamefont {T.}~\bibnamefont {Kanne}}, \bibinfo {author} {\bibfnamefont {J.}~\bibnamefont {Nyg{\aa}rd}}, \bibinfo {author} {\bibfnamefont {M.}~\bibnamefont {Kjaergaard}},\ and\ \bibinfo {author} {\bibfnamefont {V.}~\bibnamefont {Fatemi}},\ }\bibfield  {title} {\bibinfo {title} {{Andreev spin relaxation time in a shadow-evaporated InAs weak link}},\ }\bibfield  {journal} {\bibinfo  {journal} {arXiv}\ }\href {https://doi.org/10.48550/arXiv.2501.11627} {10.48550/arXiv.2501.11627} (\bibinfo {year} {2025}),\ \Eprint {https://arxiv.org/abs/2501.11627} {2501.11627} \BibitemShut {NoStop}%
\bibitem [{\citenamefont {Shvetsov}\ \emph {et~al.}(2025)\citenamefont {Shvetsov}, \citenamefont {Khola}, \citenamefont {Buccheri}, \citenamefont {Cools}, \citenamefont {Trnjanin}, \citenamefont {Kanne}, \citenamefont {Nygård},\ and\ \citenamefont {Geresdi}}]{shvetsov_2025_17117780}%
  \BibitemOpen
  \bibfield  {author} {\bibinfo {author} {\bibfnamefont {O.}~\bibnamefont {Shvetsov}}, \bibinfo {author} {\bibfnamefont {A.}~\bibnamefont {Khola}}, \bibinfo {author} {\bibfnamefont {V.}~\bibnamefont {Buccheri}}, \bibinfo {author} {\bibfnamefont {I.}~\bibnamefont {Cools}}, \bibinfo {author} {\bibfnamefont {N.}~\bibnamefont {Trnjanin}}, \bibinfo {author} {\bibfnamefont {T.}~\bibnamefont {Kanne}}, \bibinfo {author} {\bibfnamefont {J.}~\bibnamefont {Nygård}},\ and\ \bibinfo {author} {\bibfnamefont {A.}~\bibnamefont {Geresdi}},\ }\bibfield  {title} {\bibinfo {title} {Dataset for ``{A}pproaching the ultrastrong coupling regime between an {A}ndreev level and a microwave resonator"},\ }\href {https://doi.org/10.5281/zenodo.17117780} {10.5281/zenodo.17117780} (\bibinfo {year} {2025})\BibitemShut {NoStop}%
\bibitem [{\citenamefont {Annunziata}\ \emph {et~al.}(2010)\citenamefont {Annunziata}, \citenamefont {Santavicca}, \citenamefont {Frunzio}, \citenamefont {Catelani}, \citenamefont {Rooks}, \citenamefont {Frydman},\ and\ \citenamefont {Prober}}]{Annunziata2010Oct}%
  \BibitemOpen
  \bibfield  {author} {\bibinfo {author} {\bibfnamefont {A.~J.}\ \bibnamefont {Annunziata}}, \bibinfo {author} {\bibfnamefont {D.~F.}\ \bibnamefont {Santavicca}}, \bibinfo {author} {\bibfnamefont {L.}~\bibnamefont {Frunzio}}, \bibinfo {author} {\bibfnamefont {G.}~\bibnamefont {Catelani}}, \bibinfo {author} {\bibfnamefont {M.~J.}\ \bibnamefont {Rooks}}, \bibinfo {author} {\bibfnamefont {A.}~\bibnamefont {Frydman}},\ and\ \bibinfo {author} {\bibfnamefont {D.~E.}\ \bibnamefont {Prober}},\ }\bibfield  {title} {\bibinfo {title} {{Tunable superconducting nanoinductors}},\ }\href {https://doi.org/10.1088/0957-4484/21/44/445202} {\bibfield  {journal} {\bibinfo  {journal} {Nanotechnology}\ }\textbf {\bibinfo {volume} {21}},\ \bibinfo {pages} {445202} (\bibinfo {year} {2010})}\BibitemShut {NoStop}%
\bibitem [{\citenamefont {Bardeen}\ \emph {et~al.}(1957)\citenamefont {Bardeen}, \citenamefont {Cooper},\ and\ \citenamefont {Schrieffer}}]{PhysRev.108.1175}%
  \BibitemOpen
  \bibfield  {author} {\bibinfo {author} {\bibfnamefont {J.}~\bibnamefont {Bardeen}}, \bibinfo {author} {\bibfnamefont {L.~N.}\ \bibnamefont {Cooper}},\ and\ \bibinfo {author} {\bibfnamefont {J.~R.}\ \bibnamefont {Schrieffer}},\ }\bibfield  {title} {\bibinfo {title} {Theory of superconductivity},\ }\href {https://doi.org/10.1103/PhysRev.108.1175} {\bibfield  {journal} {\bibinfo  {journal} {Phys. Rev.}\ }\textbf {\bibinfo {volume} {108}},\ \bibinfo {pages} {1175} (\bibinfo {year} {1957})}\BibitemShut {NoStop}%
\bibitem [{\citenamefont {Cools}\ \emph {et~al.}(2025)\citenamefont {Cools}, \citenamefont {L{\ifmmode\acute{o}\else\'{o}\fi}pez-B{\ifmmode\acute{a}\else\'{a}\fi}ez}, \citenamefont {Buccheri}, \citenamefont {Shvetsov}, \citenamefont {Trnjanin}, \citenamefont {Hogedal},\ and\ \citenamefont {Dash}}]{Cools2025Jun}%
  \BibitemOpen
  \bibfield  {author} {\bibinfo {author} {\bibfnamefont {I.~P.~C.}\ \bibnamefont {Cools}}, \bibinfo {author} {\bibfnamefont {R.~M.}\ \bibnamefont {L{\ifmmode\acute{o}\else\'{o}\fi}pez-B{\ifmmode\acute{a}\else\'{a}\fi}ez}}, \bibinfo {author} {\bibfnamefont {V.}~\bibnamefont {Buccheri}}, \bibinfo {author} {\bibfnamefont {O.}~\bibnamefont {Shvetsov}}, \bibinfo {author} {\bibfnamefont {N.}~\bibnamefont {Trnjanin}}, \bibinfo {author} {\bibfnamefont {E.}~\bibnamefont {Hogedal}},\ and\ \bibinfo {author} {\bibfnamefont {S.~P.}\ \bibnamefont {Dash}},\ }\bibfield  {title} {\bibinfo {title} {{Losses in magnetic field resilient coplanar stripline resonators}},\ }\href {https://doi.org/10.1088/1361-6463/added5} {\bibfield  {journal} {\bibinfo  {journal} {J. Phys. D: Appl. Phys.}\ }\textbf {\bibinfo {volume} {58}},\ \bibinfo {pages} {255102} (\bibinfo {year} {2025})}\BibitemShut {NoStop}%
\bibitem [{\citenamefont {Probst}\ \emph {et~al.}(2015)\citenamefont {Probst}, \citenamefont {Song}, \citenamefont {Bushev}, \citenamefont {Ustinov},\ and\ \citenamefont {Weides}}]{10.1063/1.4907935}%
  \BibitemOpen
  \bibfield  {author} {\bibinfo {author} {\bibfnamefont {S.}~\bibnamefont {Probst}}, \bibinfo {author} {\bibfnamefont {F.~B.}\ \bibnamefont {Song}}, \bibinfo {author} {\bibfnamefont {P.~A.}\ \bibnamefont {Bushev}}, \bibinfo {author} {\bibfnamefont {A.~V.}\ \bibnamefont {Ustinov}},\ and\ \bibinfo {author} {\bibfnamefont {M.}~\bibnamefont {Weides}},\ }\bibfield  {title} {\bibinfo {title} {{Efficient and robust analysis of complex scattering data under noise in microwave resonators}},\ }\href {https://doi.org/10.1063/1.4907935} {\bibfield  {journal} {\bibinfo  {journal} {Review of Scientific Instruments}\ }\textbf {\bibinfo {volume} {86}},\ \bibinfo {pages} {024706} (\bibinfo {year} {2015})}\BibitemShut {NoStop}%
\bibitem [{\citenamefont {Kurilovich}\ \emph {et~al.}(2021)\citenamefont {Kurilovich}, \citenamefont {Kurilovich}, \citenamefont {Fatemi}, \citenamefont {Devoret},\ and\ \citenamefont {Glazman}}]{Kurilovich2021Nov}%
  \BibitemOpen
  \bibfield  {author} {\bibinfo {author} {\bibfnamefont {P.~D.}\ \bibnamefont {Kurilovich}}, \bibinfo {author} {\bibfnamefont {V.~D.}\ \bibnamefont {Kurilovich}}, \bibinfo {author} {\bibfnamefont {V.}~\bibnamefont {Fatemi}}, \bibinfo {author} {\bibfnamefont {M.~H.}\ \bibnamefont {Devoret}},\ and\ \bibinfo {author} {\bibfnamefont {L.~I.}\ \bibnamefont {Glazman}},\ }\bibfield  {title} {\bibinfo {title} {{Microwave response of an Andreev bound state}},\ }\href {https://doi.org/10.1103/PhysRevB.104.174517} {\bibfield  {journal} {\bibinfo  {journal} {Phys. Rev. B}\ }\textbf {\bibinfo {volume} {104}},\ \bibinfo {pages} {174517} (\bibinfo {year} {2021})}\BibitemShut {NoStop}%
\bibitem [{\citenamefont {Jung}\ \emph {et~al.}(2013)\citenamefont {Jung}, \citenamefont {Butz}, \citenamefont {Shitov},\ and\ \citenamefont {Ustinov}}]{Jung2013Feb}%
  \BibitemOpen
  \bibfield  {author} {\bibinfo {author} {\bibfnamefont {P.}~\bibnamefont {Jung}}, \bibinfo {author} {\bibfnamefont {S.}~\bibnamefont {Butz}}, \bibinfo {author} {\bibfnamefont {S.~V.}\ \bibnamefont {Shitov}},\ and\ \bibinfo {author} {\bibfnamefont {A.~V.}\ \bibnamefont {Ustinov}},\ }\bibfield  {title} {\bibinfo {title} {{Low-loss tunable metamaterials using superconducting circuits with Josephson junctions}},\ }\href {https://doi.org/10.1063/1.4792705} {\bibfield  {journal} {\bibinfo  {journal} {Appl. Phys. Lett.}\ }\textbf {\bibinfo {volume} {102}},\ \bibinfo {pages} {062601} (\bibinfo {year} {2013})}\BibitemShut {NoStop}%
\bibitem [{\citenamefont {Paila}\ \emph {et~al.}(2009)\citenamefont {Paila}, \citenamefont {Gunnarsson}, \citenamefont {Sarkar}, \citenamefont {Sillanp\"a\"a},\ and\ \citenamefont {Hakonen}}]{PhysRevB.80.144520}%
  \BibitemOpen
  \bibfield  {author} {\bibinfo {author} {\bibfnamefont {A.}~\bibnamefont {Paila}}, \bibinfo {author} {\bibfnamefont {D.}~\bibnamefont {Gunnarsson}}, \bibinfo {author} {\bibfnamefont {J.}~\bibnamefont {Sarkar}}, \bibinfo {author} {\bibfnamefont {M.~A.}\ \bibnamefont {Sillanp\"a\"a}},\ and\ \bibinfo {author} {\bibfnamefont {P.~J.}\ \bibnamefont {Hakonen}},\ }\bibfield  {title} {\bibinfo {title} {Current-phase relation and josephson inductance in a superconducting cooper-pair transistor},\ }\href {https://doi.org/10.1103/PhysRevB.80.144520} {\bibfield  {journal} {\bibinfo  {journal} {Phys. Rev. B}\ }\textbf {\bibinfo {volume} {80}},\ \bibinfo {pages} {144520} (\bibinfo {year} {2009})}\BibitemShut {NoStop}%
\bibitem [{\citenamefont {Petersson}\ \emph {et~al.}(2010)\citenamefont {Petersson}, \citenamefont {Petta}, \citenamefont {Lu},\ and\ \citenamefont {Gossard}}]{PhysRevLett.105.246804}%
  \BibitemOpen
  \bibfield  {author} {\bibinfo {author} {\bibfnamefont {K.~D.}\ \bibnamefont {Petersson}}, \bibinfo {author} {\bibfnamefont {J.~R.}\ \bibnamefont {Petta}}, \bibinfo {author} {\bibfnamefont {H.}~\bibnamefont {Lu}},\ and\ \bibinfo {author} {\bibfnamefont {A.~C.}\ \bibnamefont {Gossard}},\ }\bibfield  {title} {\bibinfo {title} {Quantum coherence in a one-electron semiconductor charge qubit},\ }\href {https://doi.org/10.1103/PhysRevLett.105.246804} {\bibfield  {journal} {\bibinfo  {journal} {Phys. Rev. Lett.}\ }\textbf {\bibinfo {volume} {105}},\ \bibinfo {pages} {246804} (\bibinfo {year} {2010})}\BibitemShut {NoStop}%
\end{thebibliography}%

\end{document}